\pdfoutput=1
\documentclass{Jinst}
 \let\ifpdf\relax   

\usepackage{graphicx}
\usepackage{amssymb}
\usepackage{epstopdf}
\usepackage{units}
\usepackage{amsmath}
\usepackage{amsfonts}
\usepackage{amssymb}
\usepackage{mathptmx}
\usepackage{booktabs}
\usepackage{mathtools}
\newcommand{\nuclide}[3][]{\ensuremath{\prescript{#3}{#1}{\mathrm{#2}}}}

\newlength{\figwidth}
\newlength{\figheight}
\newlength{\dualfigwidth}
\def\spac{1}
\renewcommand{\deg}{\ensuremath{^\circ}}
\providecommand{\nuebar}{\ensuremath{\overline{\nu}_e}}
\providecommand{\sinth}{\ensuremath{\sin^2 2 \theta_{13}}}
\providecommand{\tot}{\ensuremath{\theta_{13}}}
\providecommand{\dml}{\ensuremath{\Delta m^2_{31}}}

\title{Low-Background Monitoring Cameras for the Daya Bay Antineutrino Detectors} 
\author{H.R.~Band$^a$, J.J.~Cherwinka$^b$, K.M.~Heeger$^a$, P.~Hinrichs$^a$, 
  M.C.~McFarlane$^a$\thanks{Corresponding Author}~, 
  W.~Wang$^a$, D.M.~Webber$^a$, T.~Wise$^a$, Q.~Xiao$^b$\\
  \llap{$^a$} 
  Department of Physics,\\
  University of Wisconsin - Madison\\
  Madison, WI 53706, USA\\
  \llap{$^b$} 
  Physical Sciences Laboratory,\\
  University of Wisconsin - Madison\\  
  Stoughton, WI 53589, USA\\
  \llap{$^*$}E-mail: \email{mcmcfarlane@wisc.edu}
}

\abstract{
  The Daya Bay Reactor Neutrino Experiment is designed to measure the neutrino mixing angle
  $\theta_{13}$ to world-leading precision. 
  The experiment deploys identical antineutrino
  detectors at distances of \unit[400-1900]{m} from six reactors in Daya Bay, China. 
  Each detector incorporates two general-purpose monitoring cameras to 
  ensure their safe construction, transportation and operation.
  The cameras must meet usage goals while satisfying stringent constraints on radioactivity, materials compatibility, 
  interference and reliability. 
  This article describes the system design, integration, operation and 
  performance. 
}


\begin{document}

\tableofcontents

\section{The Daya Bay Experiment}
\label{sec:Experiment}

The Daya Bay experiment~\cite{Guo:2007ug} is designed to measure the neutrino mixing angle $\theta_{13}$
to world-leading precision~\cite{An:2012eh}.
By its completion, 
eight identical, 20-ton detectors~\cite{ad12:2012pc} will be distributed over three experimental halls 
to measure antineutrino fluxes from six \unit[2.95]{GW} nuclear reactors~\cite{cgnpc}. 

\begin{center}
  \begin{figure}[htbp]
    \centering
    \includegraphics[height = 0.875 \figheight]{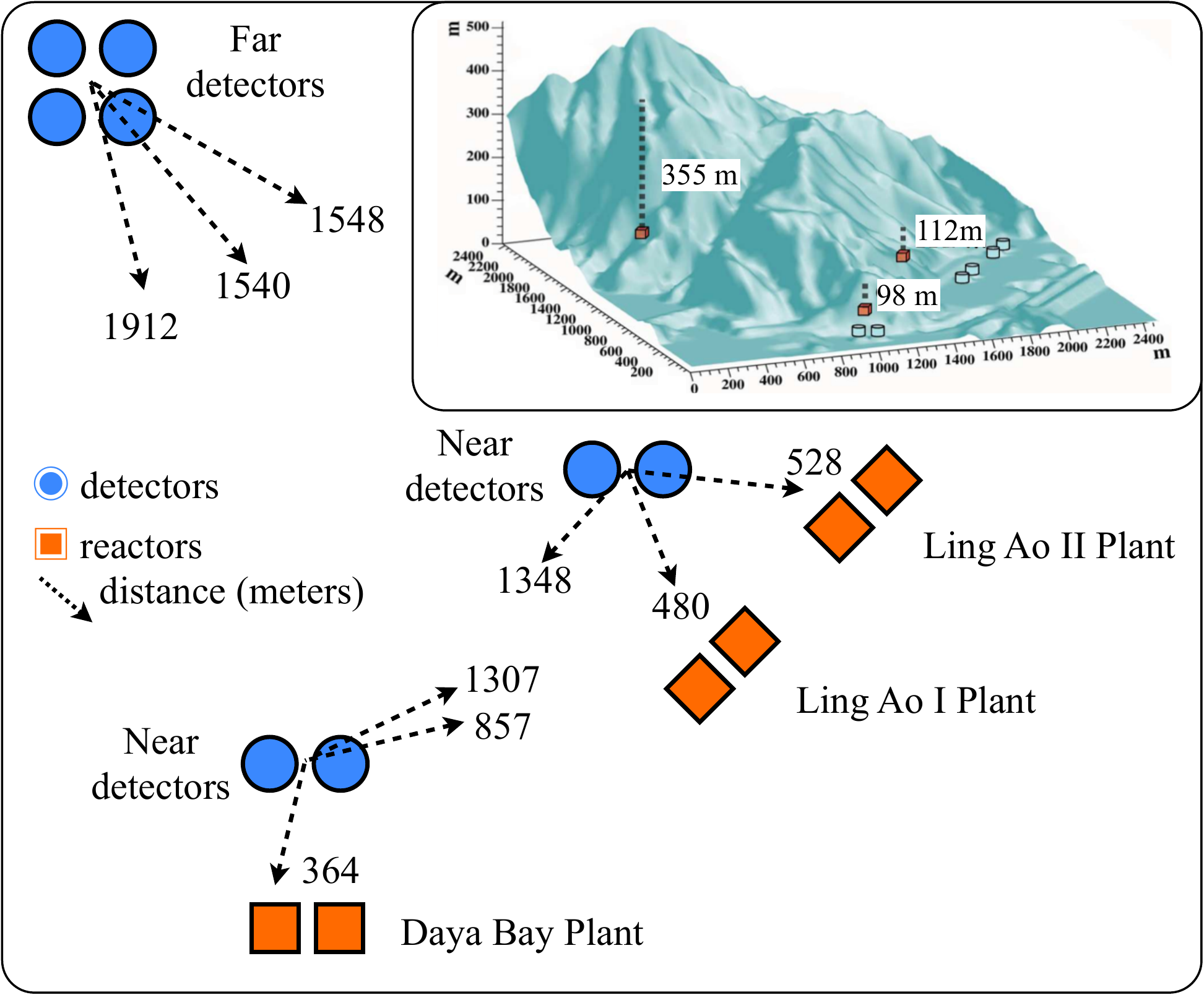}
    \includegraphics[height = 0.875 \figheight]{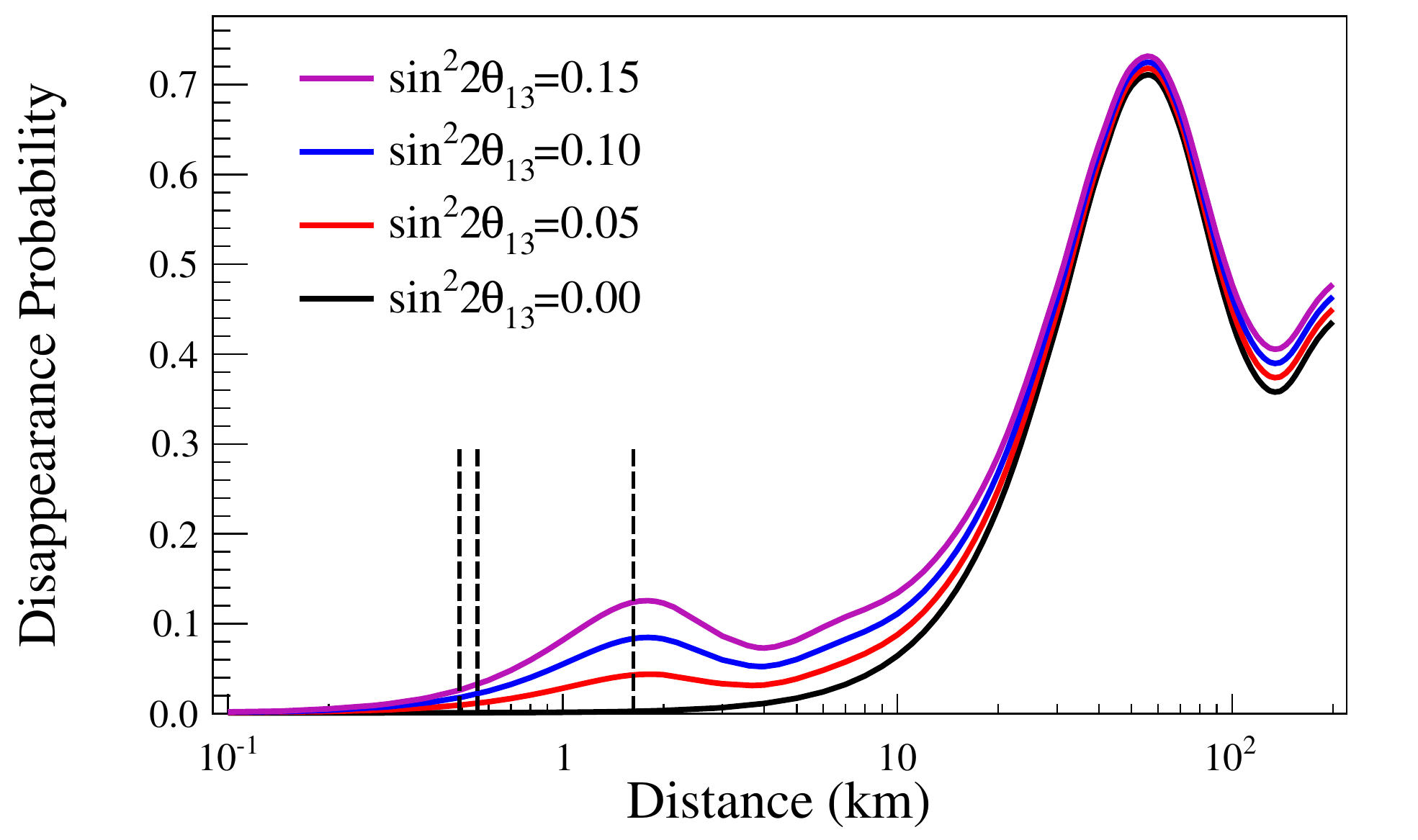}
    \caption{
      \label{fig:geo}
      Left: The topography of the Daya Bay experiment including the eight detectors at three sites,
      six reactor cores, and distances among them. The mountain overburden is shown in the inset.
      Right: The oscillation probability for reactor antineutrinos. The Daya Bay near sites
      sample antineutrinos with little probability of oscillation, while the far site
      detects them at maximal sensitivity to $\sin^2 2\theta_{13}$ and minimal sensitivity to 
      other mixing parameters. The vertical dashed lines indicate the flux-weighted 
      average baselines of the three sites.
    }
  \end{figure}
\end{center}

Nuclear reactors are a prodigious source of antineutrinos  
with energies below \unit[10]{MeV}.
For the antineutrino energies and propagation lengths at Daya Bay, the 
electron antineutrino oscillation probability is directly related to \sinth\ and
exhibits minimal dependence on most other oscillation parameters. 
Over the distances and energies at Daya Bay, the probability of a reactor electron antineutrino
remaining in the electron flavor state is approximated well by
Equation~\ref{eq:P}~\cite{PTP.28.870,Pontecorvo:1967fh,Gribov:1968kq}.
Figure~\ref{fig:geo} plots the exact 
oscillation probability as a function of distance traveled, integrated over 
the reactor antineutrino energy spectrum. 
\begin{equation}
  \label{eq:P}
  \mathrm{P}_{\overline{\nu}_e\rightarrow \overline{\nu}_e} \simeq 1 - \sinth \sin^2\left(1.267~ \frac{\dml (\mathrm{eV^2}) ~L(\mathrm{m})}{E(\mathrm{MeV})} \right)
\end{equation}

Oscillation between the Daya Bay reactors 
and the experimental sites reduces the electron antineutrino flux and distorts the 
detected spectra in proportion to the value of \sinth.
Figure~\ref{fig:geo} provides the topology of the Daya Bay experiment.
The two near detector sites, built to house two detectors each, 
are situated a flux-weighted average of about \unit[500]{m} from the reactors.
The far detector site, built for four detectors, is located an average
distance of about \unit[1.6]{km} from the reactors, near the oscillation maximum.

The strength of the Daya Bay experiment is the deployment of identical detector pairs to these near and 
far sites, with special emphasis on constructing identical near-far pairs.
As Figure~\ref{fig:geo} illustrates, the far site maximizes sensitivity to \tot\ 
while minimizing sensitivity to $\theta_{12}$,
$\theta_{23}$ and $\Delta m^2_{12}$. While the oscillation probability depends on $\Delta m^2_{32}$, 
the far site is located such that the event rates are equal over the range of possible values of $\Delta m^2_{32}$.
The near detectors make a precise measurement of the initial antineutrino flux  
with limited oscillation, 
and the powerful near-far pairing mitigates uncertainties on the reactor power 
and detection efficiency. 

\section{Daya Bay Antineutrino Detectors}
\label{sec:det}

\begin{center}
  \begin{figure}
    \begin{center}
      \center{\includegraphics[width = \figwidth]{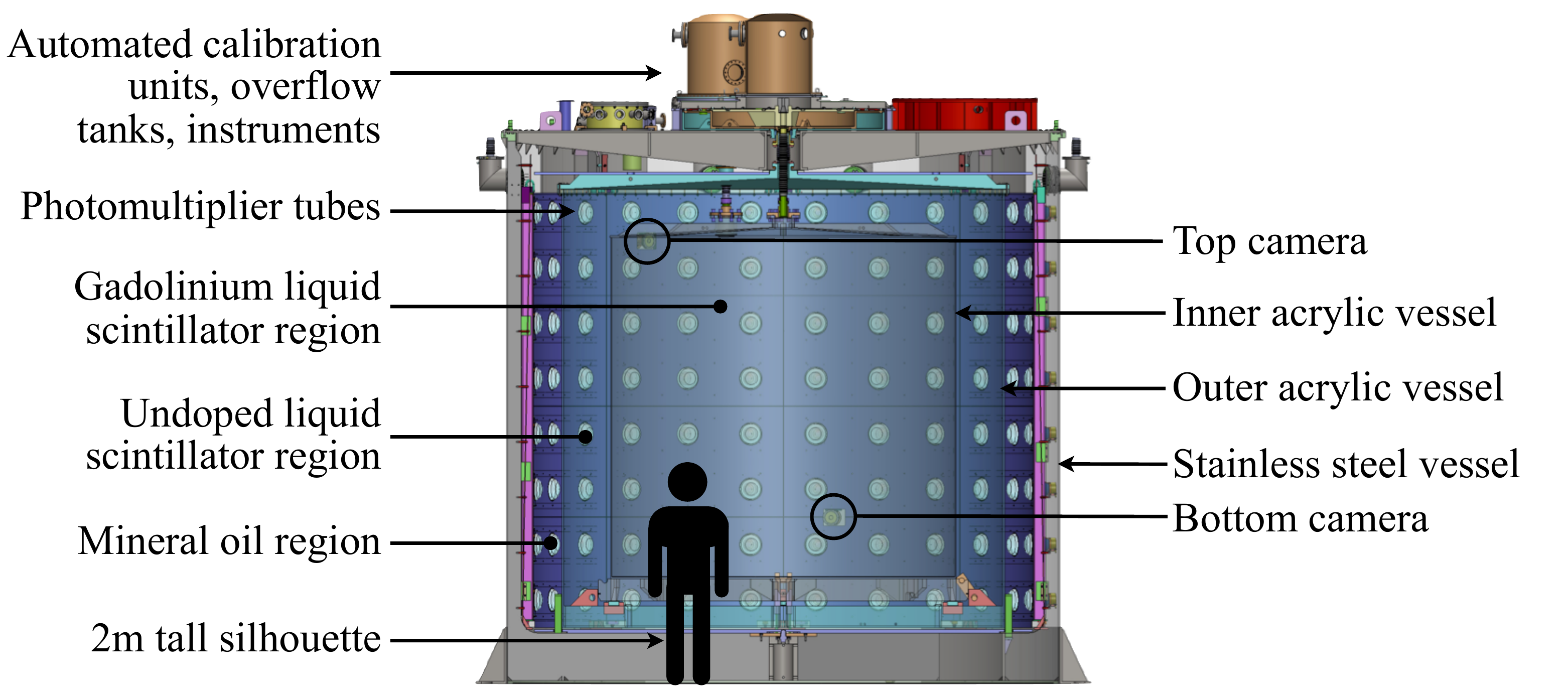}}
      \caption{\label{fig:ADcutaway}
        Cutaway view of a detector with transparent acrylic vessels. General detector components and cameras are indicated.
        The comparison silhouette is \unit[2]{m} tall.
      }
    \end{center}
  \end{figure}
\end{center}

The Daya Bay antineutrino detectors consist of a 20-ton target of gadolinium-doped liquid scintillator 
(linear alkylbenzene) surrounded by a volume of undoped liquid 
scintillator, which is in turn enveloped by a nonscintillating 
mineral oil buffer. 
These liquids are contained by two concentric acrylic vessels~\cite{Band:2012dh} enclosed in
a stainless steel vessel. The stainless steel vessel supports 192 photomultiplier 
tubes (PMTs), top and bottom reflectors, calibration units and 
multiple monitoring devices, including the cameras described in this paper. The PMTs
are distributed azimuthally around the detector in conjunction with the 
specular reflectors at the top and bottom. Figure~\ref{fig:ADcutaway} provides a labeled cross-section of 
the detector including cameras.

Antineutrino detection occurs via inverse beta decay
($\nuebar + p \rightarrow e^+ + n$) in the target region. 
An incoming antineutrino reacts with a proton, generating a neutron and a 
positron. The positron annihilates promptly into gamma rays, 
while the neutron thermalizes before capturing on a gadolinium nucleus, releasing more gammas 
after a known delay.
Thus, antineutrino events in the detectors exhibit a distinct energy and time signature, 
which allows excellent discrimination from backgrounds.
The gadolinium in this reaction was added to the target volume for its large 
neutron capture cross-section,
while the envelope of undoped scintillator functions to catch most gamma rays that escape from 
the target volume.
In this manner, fiducial cuts are unnecessary to identify antineutrino events.

Two cameras are built into each antineutrino detector 
as an integral part of their monitoring instrumentation.
These cameras provide the only visual information inside the detector,
and offer vital monitoring images during construction and transportation. 
The cameras can also cross-check the motion of 
calibration sources inside the detectors as desired.

\section{Goals and Design Requirements of the Monitoring Camera System}
\label{sec:goals}

\begin{center}
  \begin{figure}
    \begin{center}
      \center{\includegraphics[width = \figwidth]{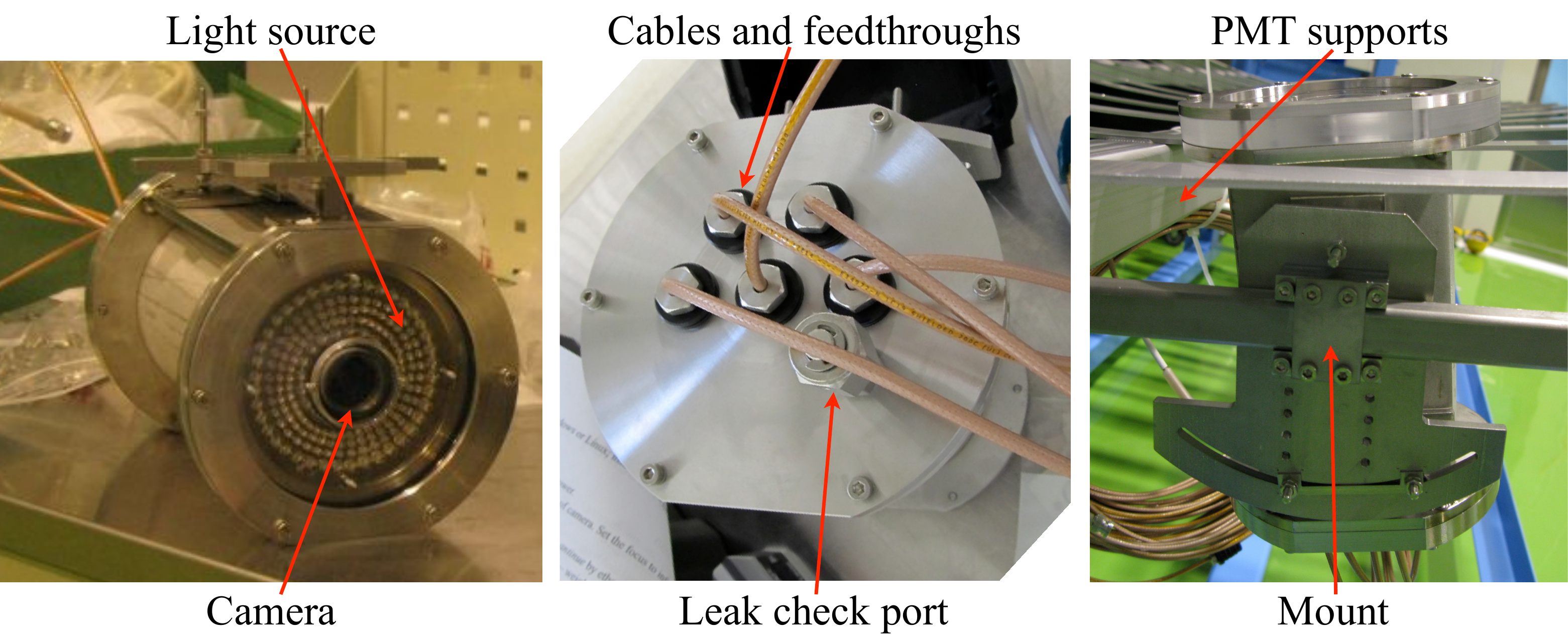}}
      \caption{\label{fig:camface}
        Labeled photographs of one camera from different vantages.
        Left: front of camera on bench during preparation for installation. 
        The camera opening and light source are prominent.
        Center: back of camera during assembly with power and readout cables, leak-tight feed-throughs and 
        leak check port visible.
        Right: camera on PMT support ladder showing the adjustable mount that
        attaches it in place. 
        The components and system design are explicated in the sections to follow.
      }
    \end{center}
  \end{figure}
\end{center}

The cameras are first shown in Figure~\ref{fig:camface}. They are designed to monitor many processes, including 
detector filling, transportation, installation and calibration, as well as to provide a flexible,
on-demand investigative tool. These processes dictate the system design goals.

One of the principal uses of the camera is the monitoring of the detector liquids during filling. 
The goal during detector filling is to maintain equal liquid levels within
\unit[5]{cm} at all times as all three fluids are pumped in simultaneously, at different speeds. 
Although there are a variety of instruments measuring the fluid mass flow, the cameras
offer the only visual verification.
The system must be capable of providing live visual confirmation of all three fluid levels inside the detector,
especially when the liquids reach the tops and bottoms of the acrylic vessels.

Similarly, transportation of the detectors, which occurs before and after filling,
is a process that requires careful monitoring to avoid damage. 
The cameras must help ensure the safety of the move by displaying the 
position of various detector elements in real time. In addition, the system must 
provide a photographic record of the positions of internal components
before and after filling, transport and installation, especially of 
the bellows that attach the inner acrylic vessels to the top of the detector, a sample of PMTs, 
the reflectors, and the connections at the bases of the acrylic vessels.

Detector calibration is another process of interest. 
The automated calibration, whereby a radioactive 
source is periodically lowered under gravity though the detector volumes,
is a well-regulated process that nonetheless benefits from the availability
of the cameras on demand. Less frequent is the manual calibration, whereby a
source is inserted along an extendible, rotatable arm from a platform atop the detector.
Either calibration procedure poses the risk of inserting an active 
source of radiation into the detector, and while the likelihood of an error is low, the consequence would be
high. The camera must be capable of cross-checking the source deployment whenever desired.

Finally, the cameras must offer an effective general-purpose inspection tool. They have to 
satisfy the explicit list of monitoring requirements while remaining as 
versatile as possible, able to provide a wide, detailed view on limited notice.

\begin{figure}[htbp]
  \centering
  \includegraphics[width=\figwidth]{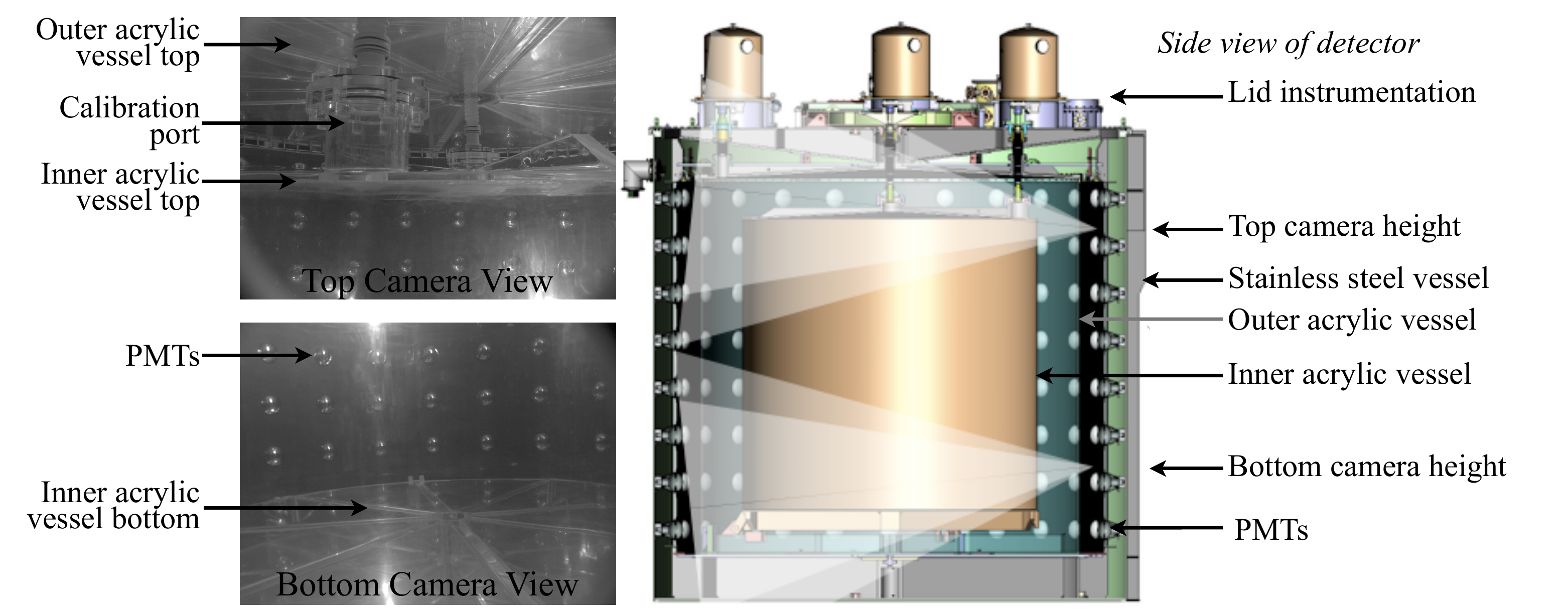}
  \caption{
    Left: Views from the cameras in an empty detector with some features labeled for reference.
    Right: Cutaway view of the detector with opaque inner acrylic vessel.
    Shaded triangles show design vertical field of view, which is 43\deg\ in an empty detector 
    and 30\deg\ when filled with liquid scintillator.
    Cameras are installed with a 6\deg\ tilt away from the center of the detector to 
    provide a better view of critical features. 
  }
  \label{fig:FOVside}
\end{figure}

\begin{figure}[htbp]
  \centering
  \includegraphics[width=\figwidth]{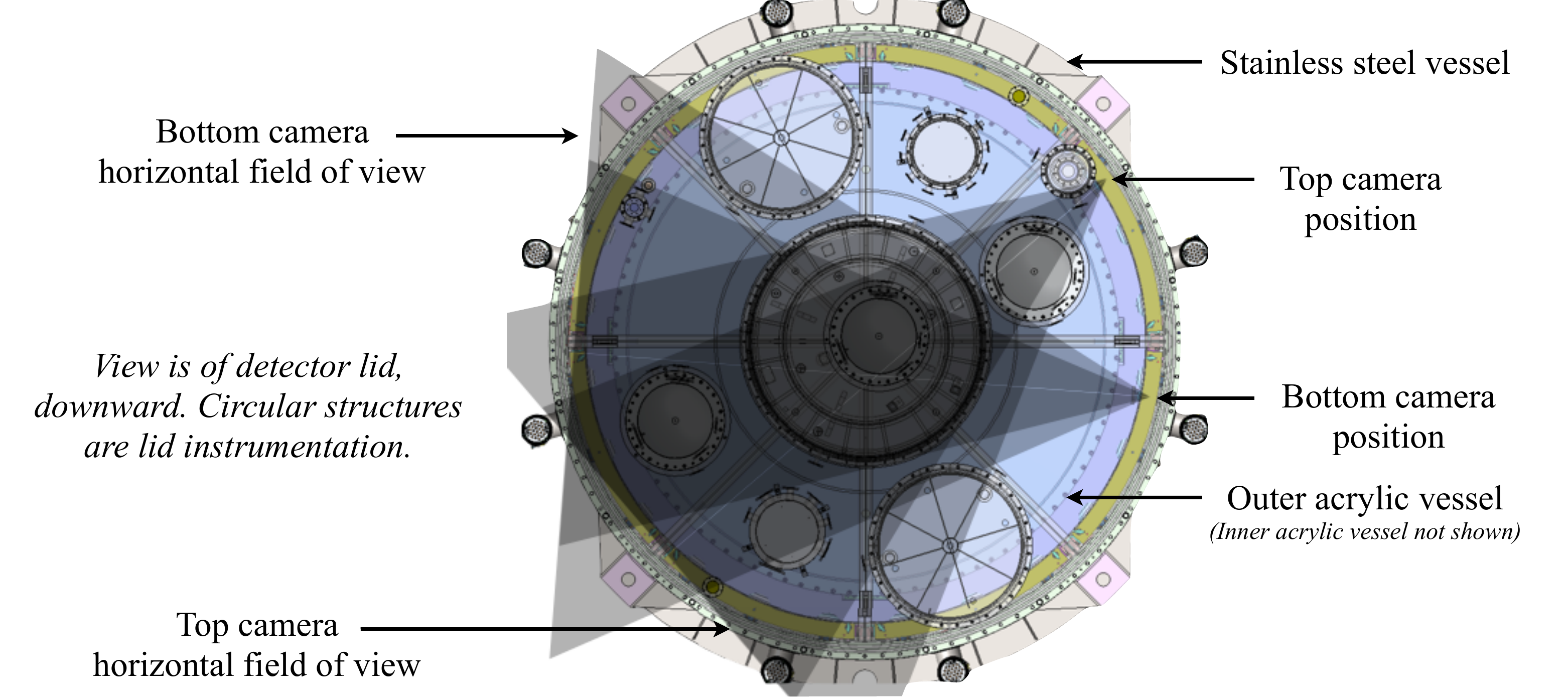}
  \caption{
    Top-down view of transparent detector lid, with camera locations and detector features indicated.
    Shaded regions show design lateral field of view: 56\deg\ in an empty detector and 37\deg\ after filling. 
    Both cameras are panned 7\deg~counterclockwise with respect to this view axis in order to photograph more critical features.
  }
  \label{fig:FOVtop}
\end{figure}

\subsection{Design Specifications} 
\label{sec:DesSpec}

The cameras' visual requirements are established by the critical features of the detector that may need to 
be photographed.
Critical features are those that are pertinent to the monitoring tasks described in Section~\ref{sec:goals}.
This list includes liquid levels during filling, 
bellows connecting the tops of the acrylic vessels with 
the lid of the detector through which calibration and filling devices pass,
the connections among the acrylic vessels, reflectors, and other interfaces among components.
From the closest bellow to the fluid levels across the detector, the depth of field required of the 
cameras is \unit[75]{cm} to \unit[450]{cm}, and
the required field of view is at least 30\deg\ horizontally and vertically. 
The design field of view is shown in Figures~\ref{fig:FOVside} and~\ref{fig:FOVtop}.
A goal of \unit[1]{mm} resolution throughout the
detector is established from the features that need to be resolved finely, such as 
bond gaps in the acrylic, or crazing that could develop. 
These requirements compete with one another and force 
compromise. For instance, a deep field generally requires a large f-stop (a narrow aperture), but the limited available light
necessitates a wide aperture. This all must be reconciled empirically. 

A clear requirement of this system is that it provide its own light, since the detector is light-tight
by necessity. Furthermore, this light must not interfere with the PMTs, or otherwise with data taking. 
As a result, the system must provide both normal white light and infrared light
well outside the PMTs' wavelength domain, and usage of both must be carefully controlled.

Equally vital, the system must meet these requirements without interfering with the
operation of the detector.
The materials chosen for the project must not introduce excessive radioactive backgrounds,
nor chemically interact with the mineral oil buffer.
The collaboration has established a radioactivity goal of \unit[50]{Hz} or less per detector
for energies greater than \unit[1]{MeV}. Any more radioactivity
risks increasing systematic uncertainties and detector dead time.
Failure to use compatible materials may cause increased opacity in mineral oil or acrylic over time.
The construction of the system therefore uses materials found elsewhere in the detector,
assembled in a clean environment such as a nitrogen bag or a class 10000 or better clean room.

Finally, the camera system must be robustly constructed to be operable throughout the five-year expected 
lifetime of a detector. Repairs are impossible once the detectors are filled, so the design 
must be simple and resilient. Mechanical motion and unnecessary electrical components are discouraged, 
data lines must be reliable, and the enclosure must be leak-tight.

\section{Design of the Monitoring Camera System}
\label{sec:design}

\begin{center}
  \begin{figure}
    \begin{center}
      \center{\includegraphics[width = \figwidth]{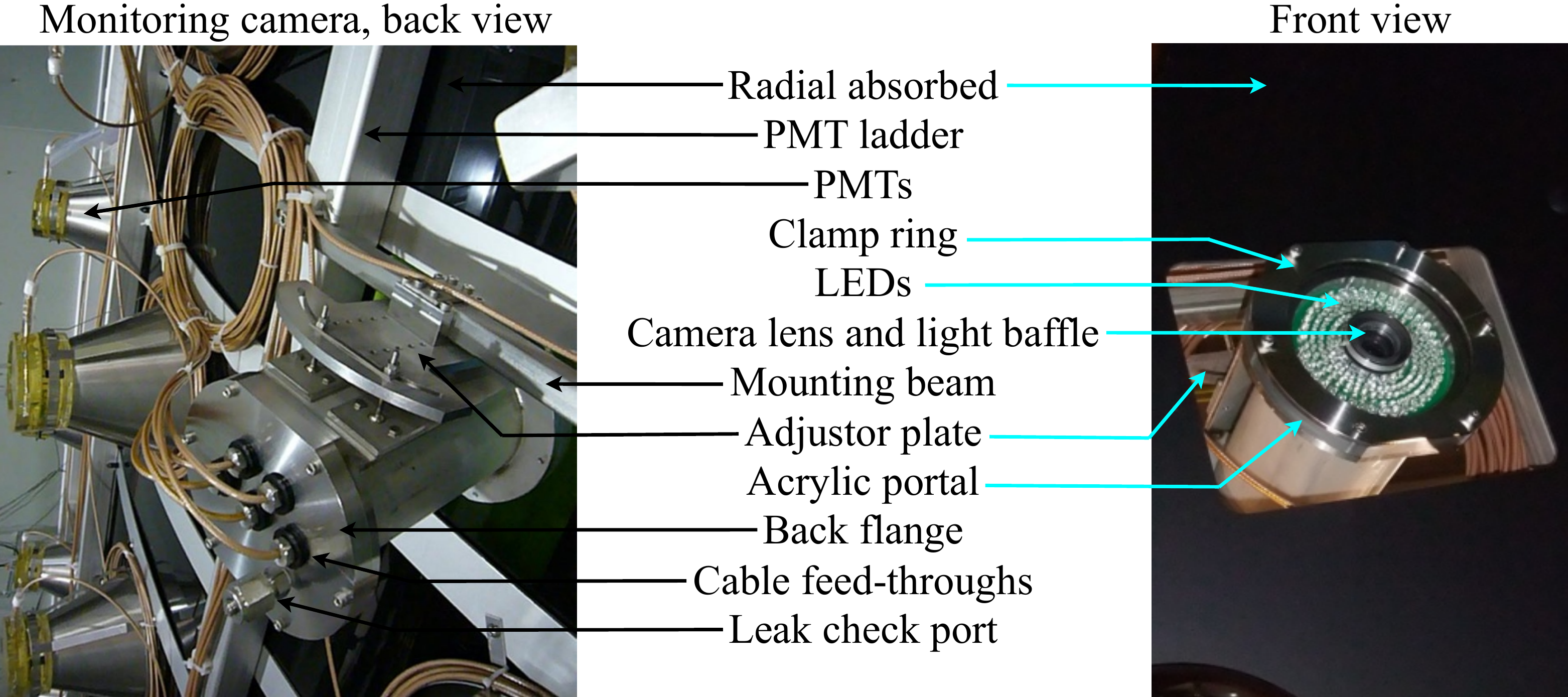}}
      \caption{\label{fig:CaminAD}
        Left:  Back view of camera installed on a PMT ladder.
        Right: Front view of camera installed on a PMT ladder.
        Major features of the camera assembly and PMT ladder are labeled. 
      }
    \end{center}
  \end{figure}
\end{center}

Figure~\ref{fig:CaminAD} shows a single camera installed on a PMT ladder, the structure of vertical and
horizontal beams that secure the PMTs. The visible components
are labeled and discussed in this section. Together, Figures~\ref{fig:ADcutaway}-\ref{fig:CaminAD} 
portray the camera system in the context of the detector.

\subsection{Camera and Optics}

\begin{figure}[h]
  \begin{center}
    \includegraphics[width = \figwidth]{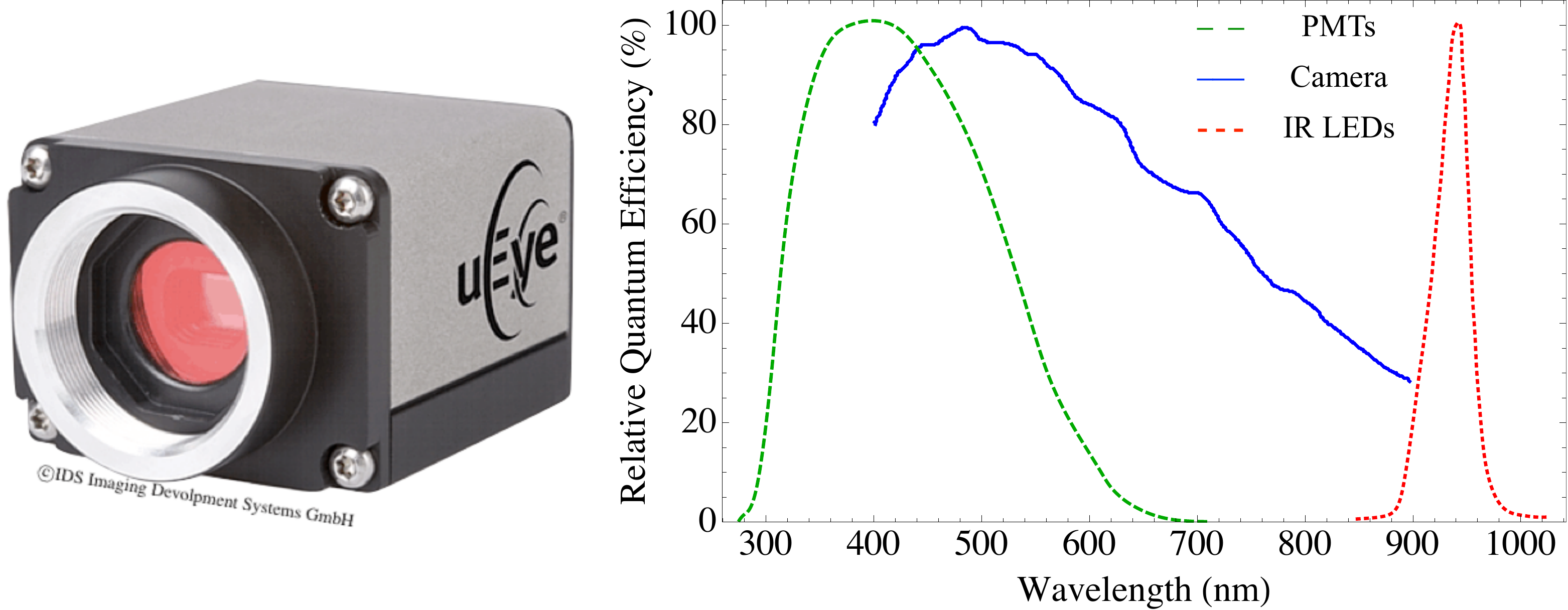}
    \caption{
      \label{fig:camera}
      Left: The IDS {\it UI5480SE-MGL}, the camera chosen for this application.
      Right: 
      The relative quantum efficiency of: the camera sensor (blue solid line)~\cite{IDS, aptina:data},
      the emission spectrum of the infrared LEDs (red dotted line)~\cite{vishay}, and
      the Hamamatsu~{\it R5912} photomultiplier tubes (PMTs) used in the detectors (green dashed line)~\cite{hamamatsu}. 
      The LED spectrum overlaps with the camera sensor spectrum, while 
      overlapping minimally with the PMTs.
    }
  \end{center}
\end{figure}

The camera chosen for this project is the small machine vision camera \textit{UI5480SE-MGL}
manufactured by IDS Imaging Development Systems~\cite{IDS}, shown in Figure~\ref{fig:camera}. 
The {\it UI5480SE-MGL} is compact and offers \unit[5]{MB} resolution, infrared sensitivity, and ethernet connectivity.
At 5 megapixels per image, the camera achieves the goal of \unit[1]{mm} resolution across the detector with a 
lens of focal length \unit[6]{mm} or longer. 
As such, the camera is not the limiting factor in resolution, but rather the dim light and the 
polish grade of the acrylic vessels. 
The camera's infrared sensitivity~\cite{aptina:data} is sufficient to photograph the detector without white light,
which Figure~\ref{fig:camera} demonstrates. 
In the \unit[900]{nm}-\unit[1000]{nm} range where the quantum efficiency is not shown in Figure~\ref{fig:camera}, 
the relative quantum efficiency is approximately half the value at \unit[900]{nm}~\cite{aptina:priv}. 
Choosing the monochrome model improves the overall sensitivity by omitting unnecessary color filters. 

Since video cameras are typically not constructed for low radioactivity measurements, they can include 
materials otherwise too radioactive for the detectors.
For instance, materials such as glass and aluminum are typically not used in low-background detectors, but 
are difficult to avoid in cameras and lenses. The solution is a camera with little mass enclosed 
in a low-radioactivity housing. At only \unit[108]{g} and \unit[40]{mm} a side, 
the {\it UI5480SE-MGL} minimizes the introduction of potentially radioactive materials
and allows the housing to be small.

Since the camera is more than \unit[50]{m} from the control electronics,
care must be taken in the choice of readout technology for robust and reliable operation.
The data are conveyed over ethernet, which is reliable and easy to integrate into the experimental halls 
at adequate bandwidth. This is superior to other options like USB and analog coaxial models.
USB camera signals would require repeaters in sensitive locations, while
cameras with analog coaxial readouts tend to offer lower resolution.

Other types of consumer and scientific cameras were considered but found unsuitable for this application.
Customized versions of commercial, consumer-grade cameras have been used in other neutrino detectors such as Borexino~\cite{Alimonti:2000xc, Back:2004qz}.
However, handheld consumer cameras generally have infrared cut filters integrated directly onto the sensor, 
are not equipped for ethernet transmission, and are not designed to run for long periods in isolation.
Pan-tilt-zoom security cameras meet these requirements, but 
rely on motors that cannot be serviced in the event of mechanical failure.
Finally, cameras for laser applications were highly sensitive to infrared but low in resolution.

The lens chosen for this system was the {\it HFA0612} from Senko Advanced Components~\cite{hfa0612}. 
The {\it HFA0612} is a megapixel-quality lens with an infrared transmission at 60\%~\cite{senko:priv}. 
It is not specially constructed for infrared focus, but testing shows that the focus is adequate. A lens that
is constructed for both megapixel quality and infrared light is much more expensive; in this application megapixel 
quality is far more important so long as infrared transmission is 
satisfactory. The reduced transmission is offset by the ability to take long exposures; all processes of interest are very slow, or static.
The  {\it HFA0612} focal length is \unit[6]{mm}, which offers both the resolution and field of view desired,
while the lens has an adjustable iris that allows the f-stop to be optimized 
experimentally. Due to the counterplay between the wide depth of field desired and the limited light,
finding the best iris diameter empirically is crucial. 
Lens types that were considered but rejected include varifocal and plastic varieties.
Varifocal lenses, for which the focal length can be adjusted electronically, have mechanical parts that risk failure. Plastic lenses, which 
would have been less radioactive, were not available at the desired resolution. 

\subsection{Lighting}
\label{sec:light}

Since Daya Bay's antineutrino detectors detect single photon signals, the detectors must 
be light-tight, and the camera system must supply its own controllable source of illumination.
This source must not interfere with the PMTs, but must provide sufficient light 
to resolve important features across the detector. It must also be regulable so that a 
user can optimize the illumination to resolve a particular feature. 
Finally, the light must not backscatter
off the acrylic into the camera, cause blinding fluorescence in the scintillator, 
nor create more heat than can be dissipated by 
the enclosure. 

\begin{figure}[htbp]
  \centering
  \includegraphics[width=\figwidth]{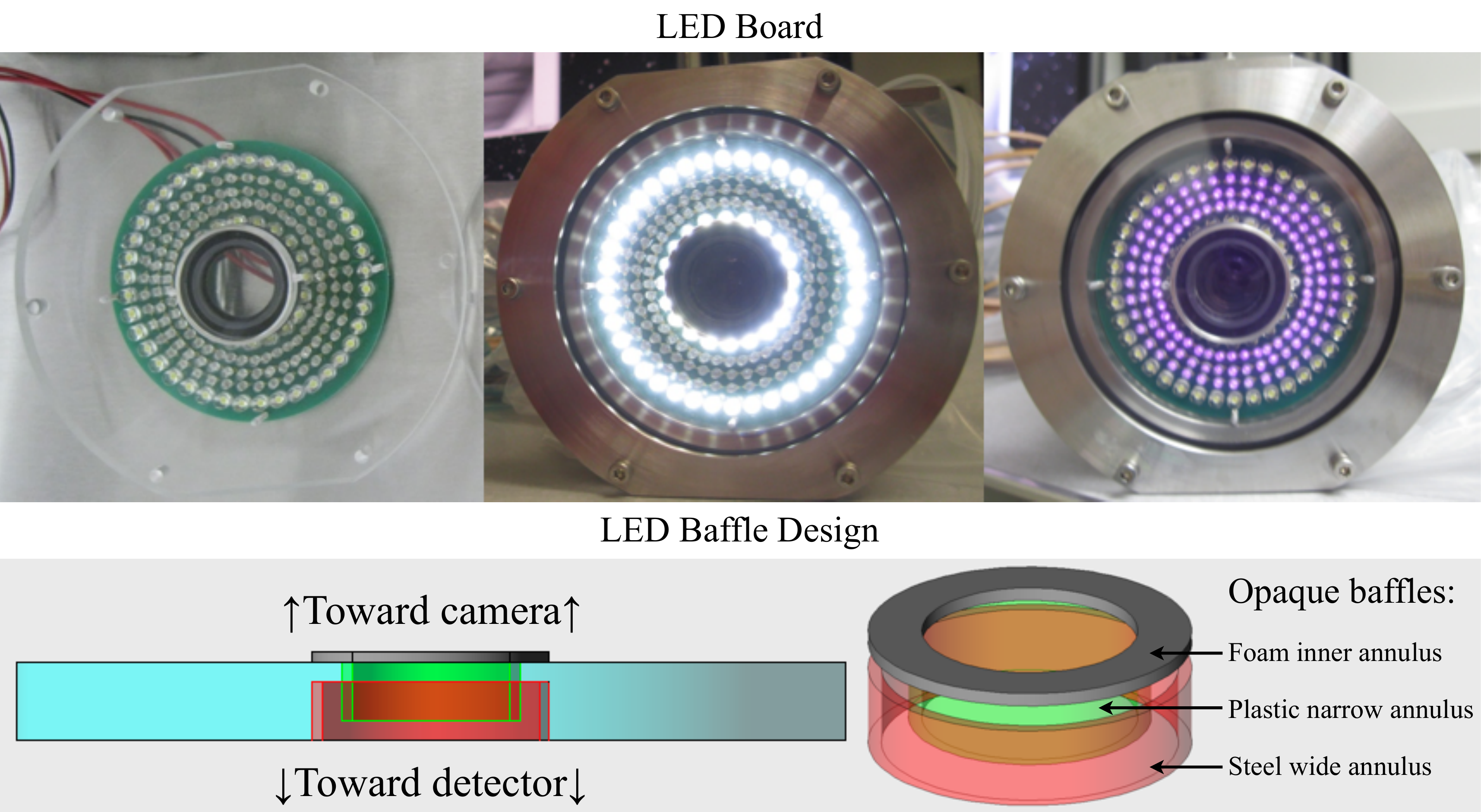}
  \caption{
    Top:
    The LED illumination element at the front of the
    camera assembly. The assembly includes five rings of LEDs, of which 
    the outer and inner rings are white LEDs and the middle three are infrared.
    The camera peers through a light baffle at the center of the portal.
    Bottom:
    Diagram of the light baffle consisting of two opposing rings and a 
    perpendicular annulus. The baffle prevents the backward
    reflection of LED light into the camera.
  }
  \label{fig:ledring}
\end{figure}

\begin{figure}[htbp]
  \centering
  \includegraphics[height=\figheight]{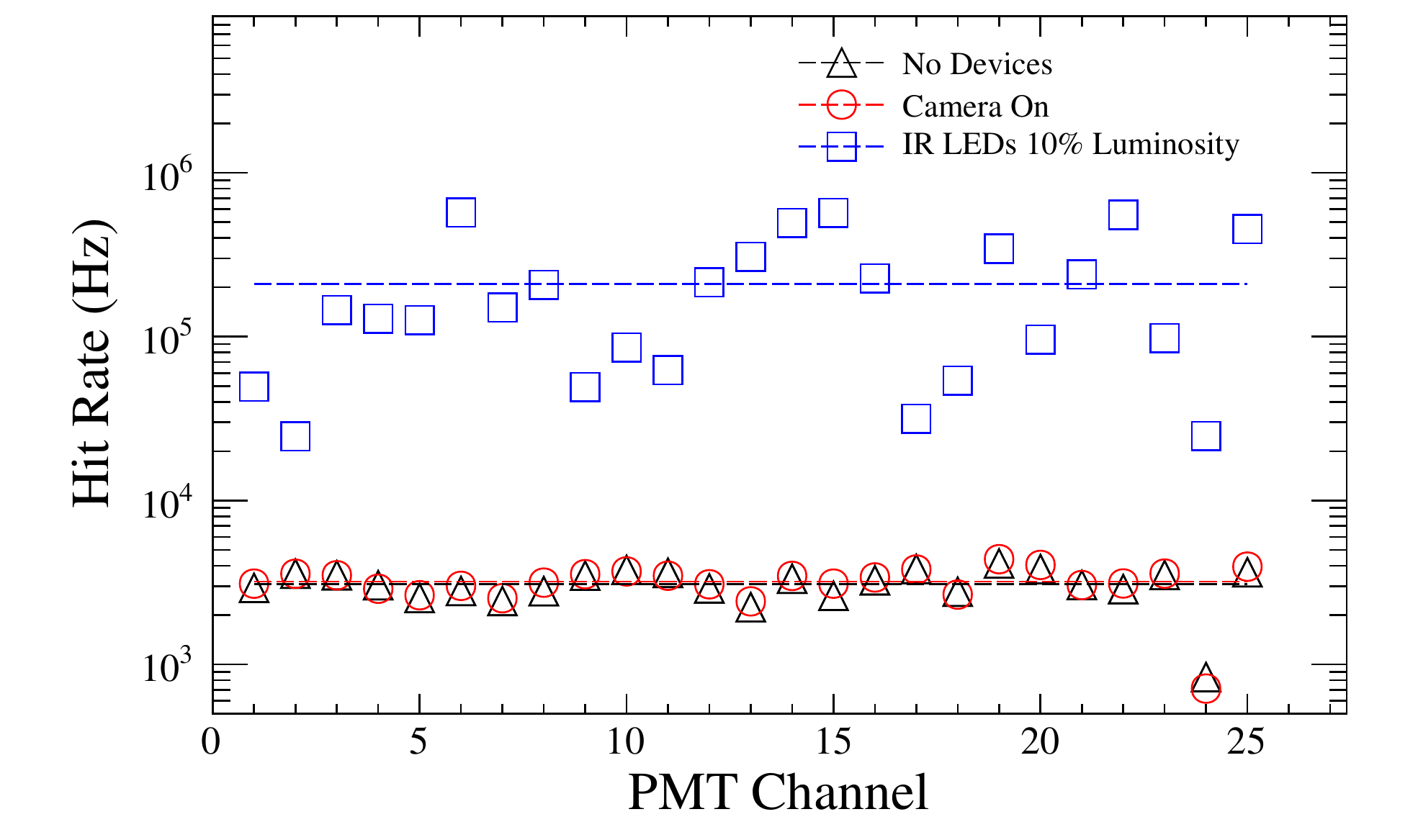}
  \caption{
    Hit rates of an empty detector with camera
    and infrared LEDs off (black triangles), camera on and infrared LEDs off 
    (red circles), and camera off and infrared LEDs at 10\% luminosity (blue squares).
    The infrared LEDs cause an increase in hit rates in a 
    dry detector. Dashed lines represent averages.
  }
  \label{fig:hitrates}
\end{figure}

In this system, the light is provided by an array of 120 infrared and 60 white LEDs
encircling the lens, shown in Figure~\ref{fig:ledring}. 
The infrared LEDs are the model {\it VSLB3940} from Vishay~\cite{vishay}, which shine
at central wavelength \unit[940]{nm}. Figure~\ref{fig:camera} juxtaposes the LED spectrum
with the quantum efficiency of the PMTs~\cite{hamamatsu} and camera.
Infrared LEDs allow the camera system to run in a manner compatible with PMTs, as required 
anytime the detector will be operated soon after illumination, or could be powered by user error. 
The consequence of running the infrared LEDs when the PMTs 
are powered is minimal because the PMT response drops sharply at about \unit[700]{nm}. 
The hit rates will increase, as shown in Figure~\ref{fig:hitrates}, 
but will return to their normal baseline quickly. 
Data is not taken when the infrared LEDs are on,
and the data acquisition system is suspended out of precaution.
It is worth noting that dark box tests of infrared LEDs and
PMTs differ from tests in the detector. The dark box PMTs 
do not show this increase in hit rates from infrared light, so it is suspected that the hit
rate increase is influenced by surrounding material.

The effect of white light on PMTs is understandably greater. When the PMTs are unpowered, illumination with the
white LEDs is preferable.
White light shone on unpowered PMTs still causes an increase in dark rates for a period of time
less than \unit[24]{hours} afterward, but no long-term effects. 
The 60 white {\it C513A-WSN} LEDs from CREE~\cite{cree} are brighter, more efficient
and illuminate within the optimal quantum efficiency  
of the camera and lens. They therefore generate less heat than the infrared LEDs.
At maximum luminosity, the 60 white LEDs draw \unit[5]{W} of power while the
infrared LEDs draw \unit[16]{W}.

The white LEDs easily illuminate the detector. The attenuation 
length in the visible-near infrared range of each optical medium is longer than \unit[10]{m},
and the liquid scintillator only absorbs in the near ultraviolet range.
Thus the liquids and acrylic neither obstruct much light nor fluoresce back into the camera.
The indices of refraction off the acrylic vessels and liquids 
are constant from the near ultraviolet to the near infrared,
so the passage of light through the acrylic and scintillating media is unimpeded 

The LEDs shine through a specially designed portal that passes light without any 
reflections back into the lens, shown in Figure~\ref{fig:ledring}. 
When the detector is unfilled, backward reflections of the LED light of the enclosure's
acrylic portal would decrease the cameras' effectiveness. To prevent glare, two 
baffles are embedded into the acrylic portal, and the lens is pressed into a black foam
annulus perpendicular to them. The arrangement successfully obstructs all glare from the lens.

Finally, both white and infrared LEDs are interlocked in software to prevent them from being activated
when the PMT high voltage is on, and the LED power cables are unplugged and locked 
out when not in use. This is discussed in Section~\ref{sec:readout}.

\subsection{Integration into Detector}

\begin{figure}[htbp]
  \centering
  \includegraphics[width=\figwidth]{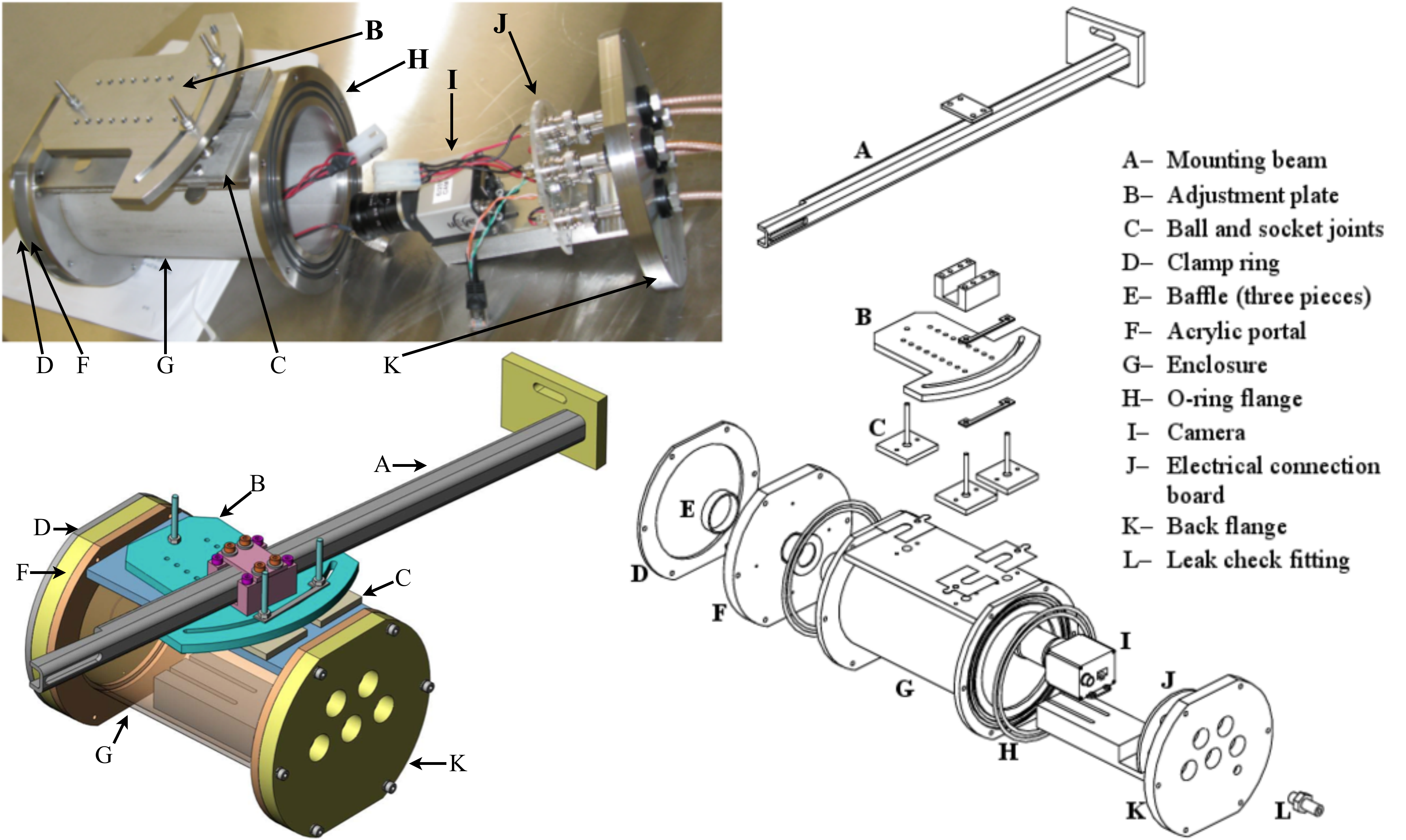}
  \caption{
    Right:
    Assembly of camera components. 
    Top Left: A camera during assembly. The front portal and LEDs have been attached; the 
    cables have been connected, passed through the back flange and secured; the
    adjustor plate has been connected and positioned and the camera awaits connection with internal 
    electrical components, followed by final closure.
    Bottom Left: Rendered model showing the camera enclosure assembled and adjusted,
    as though installed in a detector. 
  }
  \label{fig:exploded}
\end{figure}

The camera housings mount onto the PMT ladders that supports the detectors' PMTs, black
absorbing shields and some sensors. The housings are fixed within an 
adjustable, stainless steel mount that permits three translations 
and two rotations. At the front of the enclosure is an
acrylic portal through which the camera and LEDs face the detector.
At the back, a leak-tight 
flange conducts five cables. All materials that contact the mineral oil buffer
in which the system resides are either stainless steel, acrylic, nylon, viton, or Teflon: 
compatible materials used elsewhere in the detectors. The enclosure and mounting are illustrated
in Figure~\ref{fig:exploded}.

The front portal allows clear imaging and illumination of the detector while withstanding its
fluid pressure. The half-inch thick acrylic presses against two o-rings and
the front flange of the enclosure body. A clamp ring in front of the acrylic maintains
pressure on the o-rings without damaging the acrylic. The portal is not a perfect circle:
two short chords are cut off the  
portal, back flange, and o-ring flanges to improve clearance for installation.
The two concentric, opaque light baffles described in  Section~\ref{sec:light} sit tightly in
opposing grooves cut 3/8'' deep and 1/16'' wide.

The camera itself is positioned against the portal along a rod that attaches to the back flange.
The rod offers adjustability of the camera position and effectively dissipates heat into the mineral oil buffer and PMT ladder.
The back flange attaches to the enclosure like the portal, and passes
five coaxial cables that conduct electrical power and data to and from
the camera and LEDs, as discussed in Section~\ref{sec:readout}.

The cables are sealed by a feed-through plug that is identical to that used for the PMTs. 
Each cable is produced with captured seal plugs on both ends. 
Within each seal plug, a hollow bolt compresses two o-rings against
the cable and inner seal plug. 
Two more o-rings compress against the seal plug and matching hole in the flange
when the plugs are pushed into place and captured by a retaining ring. 
Overall, the enclosure contains the camera, LEDs and electronics in a leak-tight volume. 

The enclosure attaches to the PMT ladder by way of an adjustable mount that allows several degrees of freedom.
The mount and enclosure are connected by three ball and socket joints formed from round indentations on a 
flat plate welded onto the enclosure. Three threaded rod-ball bearing assemblies sit in these
indentations, held by three smaller capture plates and shims,
with the threaded rods extending through an adjustor plate. The capture plates and shims hold the threaded rods fast, 
but allow slight movement of the joints to prevent tension. The adjustor plate has one clearance hole
serving as a pivot point, and one continuous arc. The arc permits the camera to pan, while nuts on the threaded rods 
allow independent pitch and height adjustment.
Once the position is set, pairs of nuts lock the configuration into place.

\begin{figure}[htbp]
  \begin{center}
    \includegraphics[width=\figwidth]{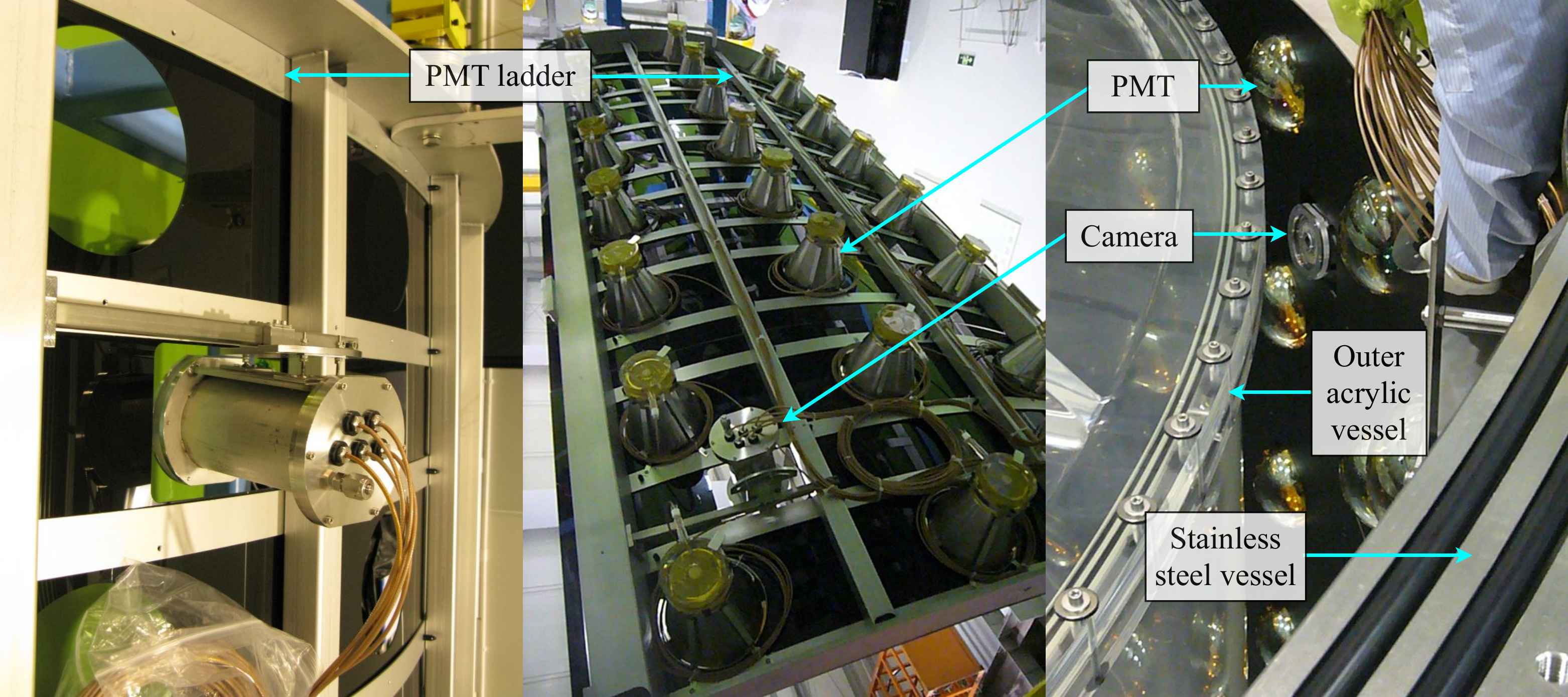} 
     \caption{
      Left:
      Back view of top camera newly installed on a PMT ladder, awaiting installation of PMTs.
      Center:
      Back view of top camera installed at the top of a PMT ladder loaded with PMTs, awaiting installation into a detector.
      Right:
      Front view of top camera installed in a detector. 
      Key detector features are labeled.
      \label{fig:openvessel}
    }
  \end{center}
\end{figure}

The cameras are installed on adjacent ladders in voids between rows of PMTs, one just below
the top row, the other above the second row from the bottom. These positions are indicated 
in the detector diagrams of Figures~\ref{fig:ADcutaway}, Figures~\ref{fig:FOVside} and \ref{fig:FOVtop}
The positions and orientations of the cameras are chosen to maximize the 
usefulness of the view. The top camera is pitched upward 6\deg\ and panned 7\deg~counterclockwise 
relative to a vertical downward axis, to visualize the most critical features,
such as the bellows and liquid levels at the tops of the acrylic vessels. The bottom camera is pitched downward to view
more low lying features of the detector such as the bottom reflector and acrylic vessel connections. 
The cameras are shown at various stages of installation in Figure~\ref{fig:openvessel}.

\subsection{System Readout and Control}
\label{sec:readout}

The control system is designed amid constraints imposed by the detectors and other electronics, particularly regarding 
materials compatibility, leak-tightness, electronic noise and inaccessibility.  
The camera requires an ethernet data connection and direct current power, while the LEDs also require direct
current power. Typical cables for power and ethernet are barred from the detectors due to materials compatibility, so
the system uses BNC coaxial cables identical to those of the PMTs for these purposes. These cables are Teflon coated
and form leak-tight seals with the existing Daya Bay seal plugs. Many non-coaxial cables, especially ethernet,
lack the uniform surface required for this seal to be leak-tight. 

\begin{figure}[htbp]
  \centering
  \includegraphics[width= \figwidth]{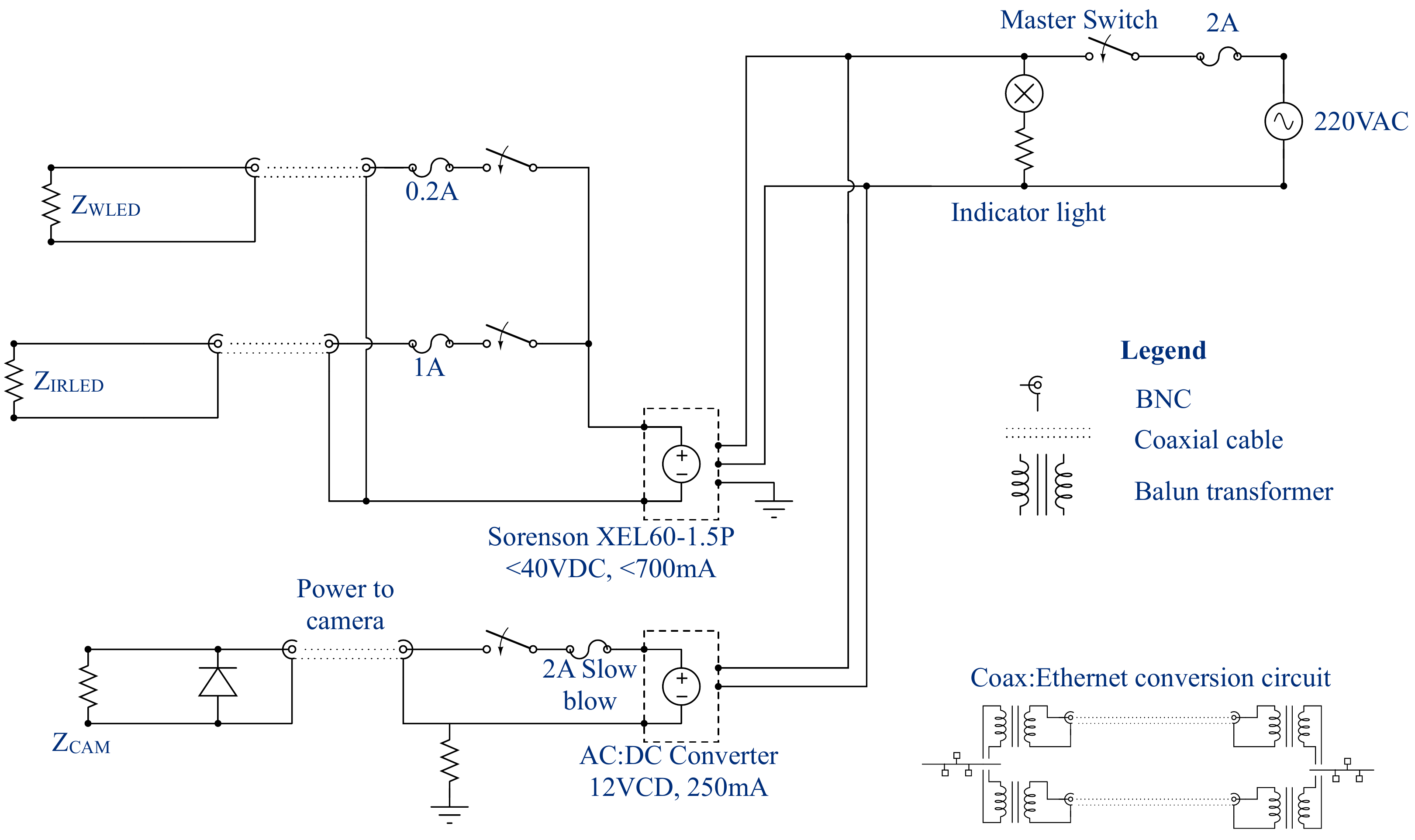}
  \caption{
    Schematic of the readout system. The center and top left of the diagram describes the readout
    system in the electronics rack outside the detector. 
    To the left, separated by a double dotted line (coaxial cable) are the 
    loads: the camera and LEDs inside the detector. To the bottom right is an ethernet to 
    coaxial cable conversion circuit using balun transformers.
    This simplified schematic depicts one camera control; 
    each readout box controls four cameras in parallel.
    An ethernet-controlled relay selects one camera and light source to power at a time. 
  }
  \label{fig:readoutbox}
\end{figure}

Signals sent to and from the camera are converted between ethernet and coaxial cables by way of balun
transformers. The baluns have a 2:1 impedance ratio to match the \unit[50]{$\Omega$} coaxial cable to the \unit[100]{$\Omega$} ethernet. 
Inside the enclosure two baluns convert two ethernet twisted pairs to two coaxial cables, 
while another pair inside the readout box reverses the conversion (Figure~\ref{fig:readoutbox}). 
A balun creates an unbalanced signal 
from the balanced, positive and negative pulses along the ethernet. While the unbalanced signal
is inherently more susceptible to noise, the coaxial cable is sufficiently shielded.
The bandwidth is reduced from the \unit[1]{Gbps} maximum of the camera to 
\unit[100]{Mbps}, but this is more than sufficient
to operate the camera within demands. Since the subjects of the camera are very slow 
processes or stationary objects, and since the camera operates in a low light environment, the frame rate 
is set as low as permitted-- \unit[0.29]{fps}-- and the bandwidth used is far lower than \unit[100]{Mbps}.

In total there are five coaxial cables: three conducting power to the camera, white and infrared LEDs, 
and two transmitting data to and from the camera.
Alternatives for fewer cables were considered and rejected for fear of superfluous components inside
the detector, which increase the risk of radioactive backgrounds, heat production, and inaccessible component failures.

In addition to converting the signal to ethernet, the readout system 
contains an ethernet-controlled relay bank and power supplies managing a total of four cameras.
DC power to all cameras and LEDs passes through the 16-switch relay bank, model {\it NER165PROXR}~\cite{ncd}
from National Control Devices, which selects one camera
and one light source to operate. 
The power supply for the LEDs is a single programmable Sorenson {\it XEL~60-1.5P}~\cite{sorenson},
and the power source for the cameras and relay bank are AC:DC converters provided by the manufacturers.
The readout box also contains an array of fuses and indicator lights for safety. 
The relay bank and power supply were chosen for their ethernet connectivity, programmability, and ease of integration into the 
control software. 

The operating software plays an important role in both running the cameras and protecting the detector and
camera system.
The LabView-based~\cite{ni} interface operates the system by controlling the relay card, power supply and camera
over the network. The power supply and relay are represented in software as Virtual Instrument Software Architecture (VISA) resources.
Custom code using drivers provided by the devices' manufacturers is integrated into the 
experiment's comprehensive control framework. 

The user is permitted to operate one of four cameras in a pair of detectors at a time, 
within the built-in safety constraints. The system is configured to preclude 
the operation of the wrong camera, and will not summon power to LEDs outside acceptable 
voltage and current ranges. Meanwhile, if the PMT high voltage is active, 
the LEDs are interlocked. Also, since the infrared LEDs
generate \unit[16]{W} of power pressed 
against the poorly dissipating acrylic surface, the user is time-limited in their usage.
Otherwise, the user has control over LED brightness and camera gain, and can select one camera out of
four connected with either type of LED at either position in the same detector. 
Only one LED strand is permitted power at a time to prevent too large a current through the system.
For best results, 
when the detector is dry, the user relies on the LEDs from the opposite camera position, taking advantage of 
the light scattering about the acrylic and reflectors. After filling, the user should illuminate the LEDs
of the same camera. The user can capture images manually or automatically at a selected rate, and the images are
immediately uploaded to a central database.

\begin{figure}[htbp]
  \centering
  \includegraphics[height = \figheight]{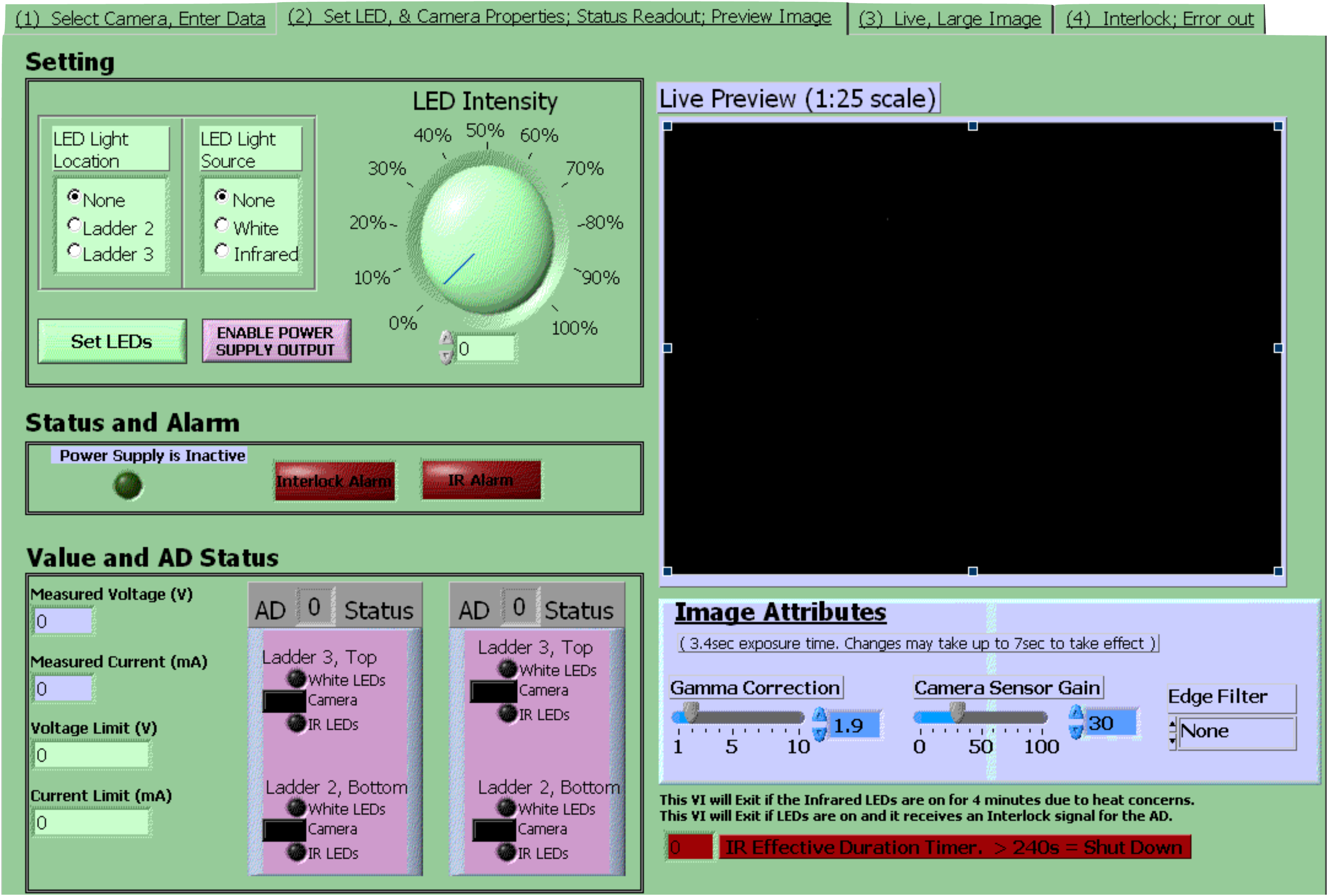}
  \caption{
    One panel of the control software. The user can operate the LED light source in the upper left,
    readout power supply and relay bank statuses in the lower left, adjust camera settings in the
    lower right, and check a preview of the image in the upper right. Functions not shown include:
    device connection, network monitors, image saving, image autosaving, image commenting, interlocks
    and some alarms.
  }
  \label{fig:VIshot}
\end{figure}

\section{Background Levels}
\label{sec:rad}

\begin{figure}[htbp]
  \centering
  \includegraphics[width = 0.9 \textwidth]{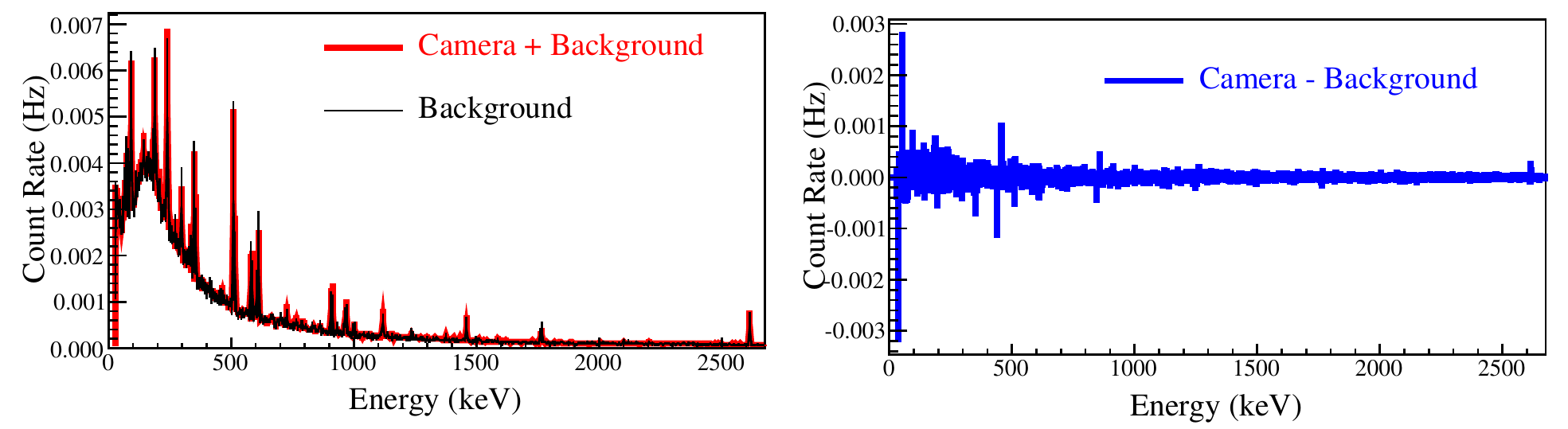}
  \caption{
    Left: Gamma spectrum of one assembled camera. Black, thin peaks denote the 
    measurement chamber alone; thick red denotes the measurement of the camera 
    system and chamber.
    Right: Difference between camera and background radiation. The camera activity is typically
    consistent with background.
    The sharp, upward and downward fluctuations in adjacent energy bins near radioactivity peaks 
    result from a small energy calibration difference between the source and background measurements.
  }
  \label{fig:rad}
\end{figure}

The radioactivity of one complete camera assembly was measured, and the 
activity from most isotopes is consistent with zero (Table~\ref{tab:rad}, Figure~\ref{fig:rad}). 
Those isotopes with nonzero activity  
are members of the $^{232}$Th decay chain, primarily $^{228}$Ac. 
In total, one camera assembly has a similar level of activity to one Hamamatsu~{\it R5912} PMT,
and the largest sources of activity are presumed to be the
glass in the lens and circuit board, and the aluminum comprising the camera case.

\begin{table}[h]
  \centering
  \begin{tabular}{llrclclc}\toprule
    {\itshape Chain}&{\itshape Nuclide}&\multicolumn{1}{c}{\itshape Line}&&\multicolumn{3}{c}{\itshape Camera Activity} & {\itshape PMT Activity}\\
    &&\multicolumn{1}{c}{(keV)}&&\multicolumn{3}{c}{(Bq)}&(Bq)\\
    \midrule
    &\phantom{$^0$}\nuclide{K}{40}&&&  0.66    & $\pm$ &   2.05   & 3.9 \\
    \midrule
    \nuclide{Th}{232}
    &$^{228}$Ac  &   911.07   &&   2.5     & $\pm$ &   0.77   &     \\
    &            &   968.97   &&   0.737   & $\pm$ &   1.13   &     \\\addlinespace[0.55ex]
    &$^{212}$Pb  &   238.63   &&   0.825   & $\pm$ &   0.43   &     \\\addlinespace[0.55ex]
    &$^{208}$Tl  &   583.14   &&   0.0957  & $\pm$ &   0.216  &     \\
    &            &  2614.47   &&   0.408   & $\pm$ &   0.315  &     \\\addlinespace[1.30ex]
    &{Total}     &            &&   0.912   & $\pm$ &   0.57   & 0.6 \\
    \midrule
    \nuclide{U}{238}
    &$^{214}$Bi  &   608.81   &&           & $ < $ &   0.45   &     \\
    &            &  1764.24   &&           & $ < $ &   1.37   &     \\\addlinespace[0.55ex]
    &$^{214}$Pb  &   294.83   &&           & $ < $ &   0.82   &     \\
    &            &   352.59   &&           & $ < $ &   0.52   &     \\\addlinespace[1.30ex]
    &Total       &            &&           & $ < $ &   0.79   & 1.9 \\
    \bottomrule
  \end{tabular}
  \caption{\label{tab:rad}
    Total radioactivity level of one camera assembly.
    The activity of an assembly is dominated by \nuclide{Th}{232}-chain decays, 
    and the total activity is lower than a typical PMT.
  }
\end{table}

\section{Performance}
\label{sec:usage}

\begin{figure}[htbp]
  \centering
  \includegraphics[width=\figwidth]{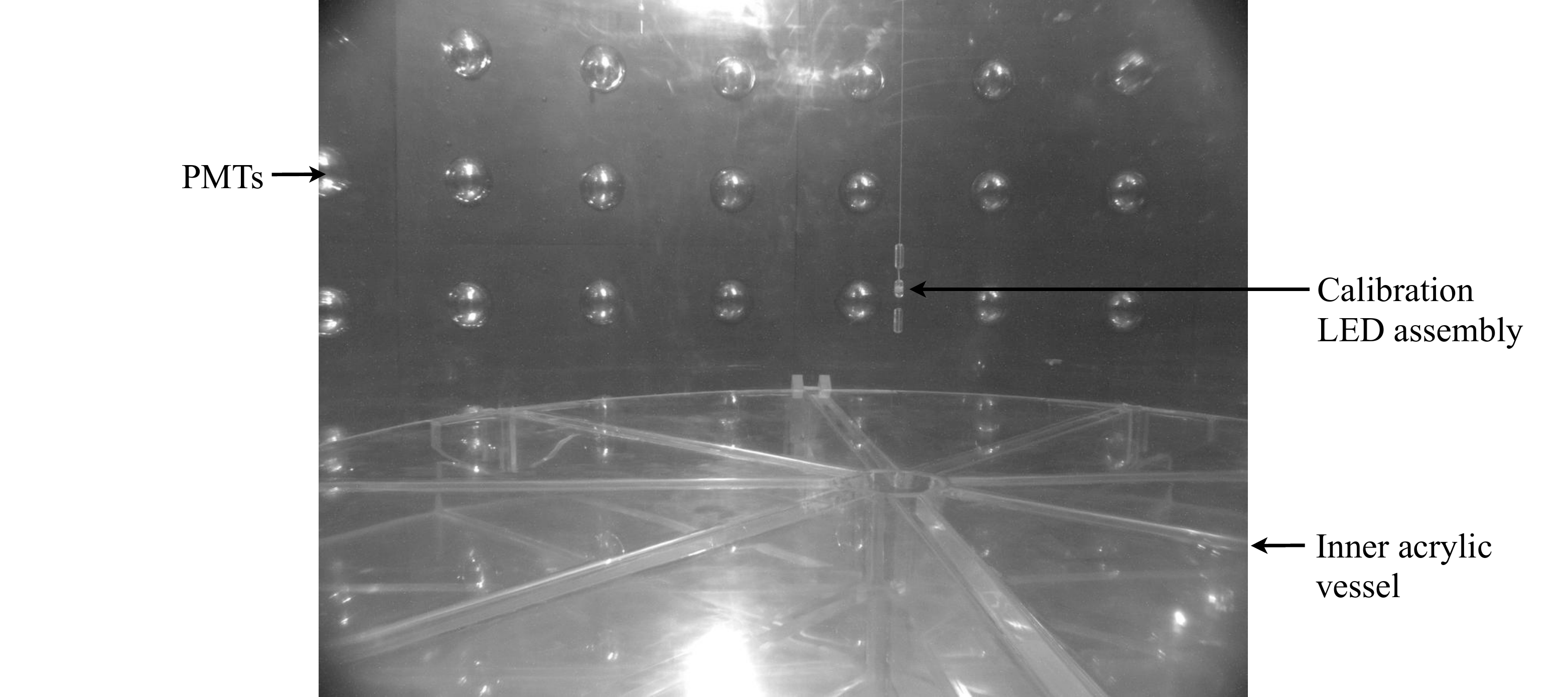}
  \caption{
    Photograph of the deployment of a calibration LED photographed 
    by the bottom camera. The detector has not been filled yet.
    The light source is the white LEDs of the top camera.
  }
  \label{fig:ACU}
\end{figure}

\begin{figure}[htbp]
  \centering
  \includegraphics[width=\figwidth]{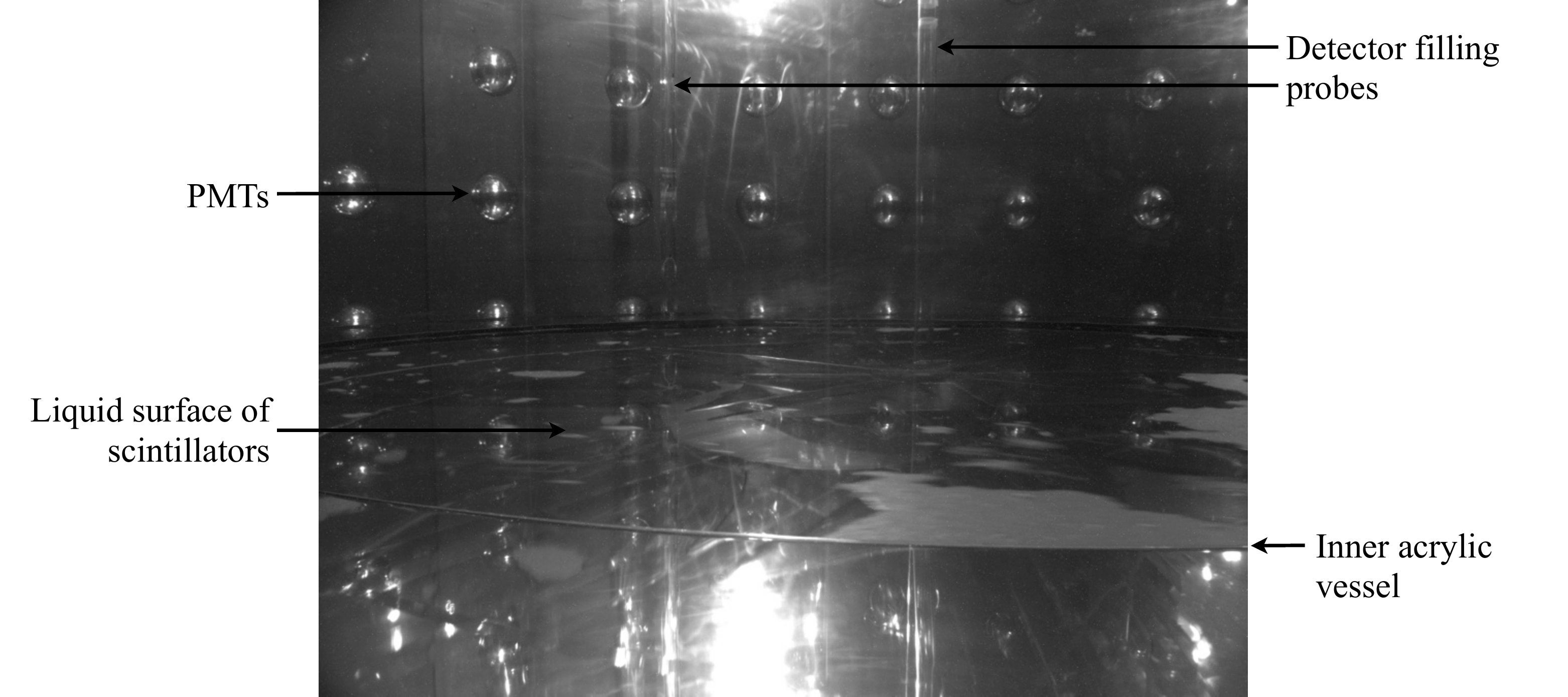}
  \caption{
    Photograph from the bottom camera during filling.
    Gadolinium-doped liquid scintillator is visible in the center of the detector,
    while undoped scintillator is visible at the bottom. The boundary between them 
    is formed by the inner acrylic vessel.
    The light source is the white LEDs of the top camera.
  }
  \label{fig:midfill}
\end{figure}

\begin{figure}[htbp]
  \centering
  \includegraphics[width=\figwidth]{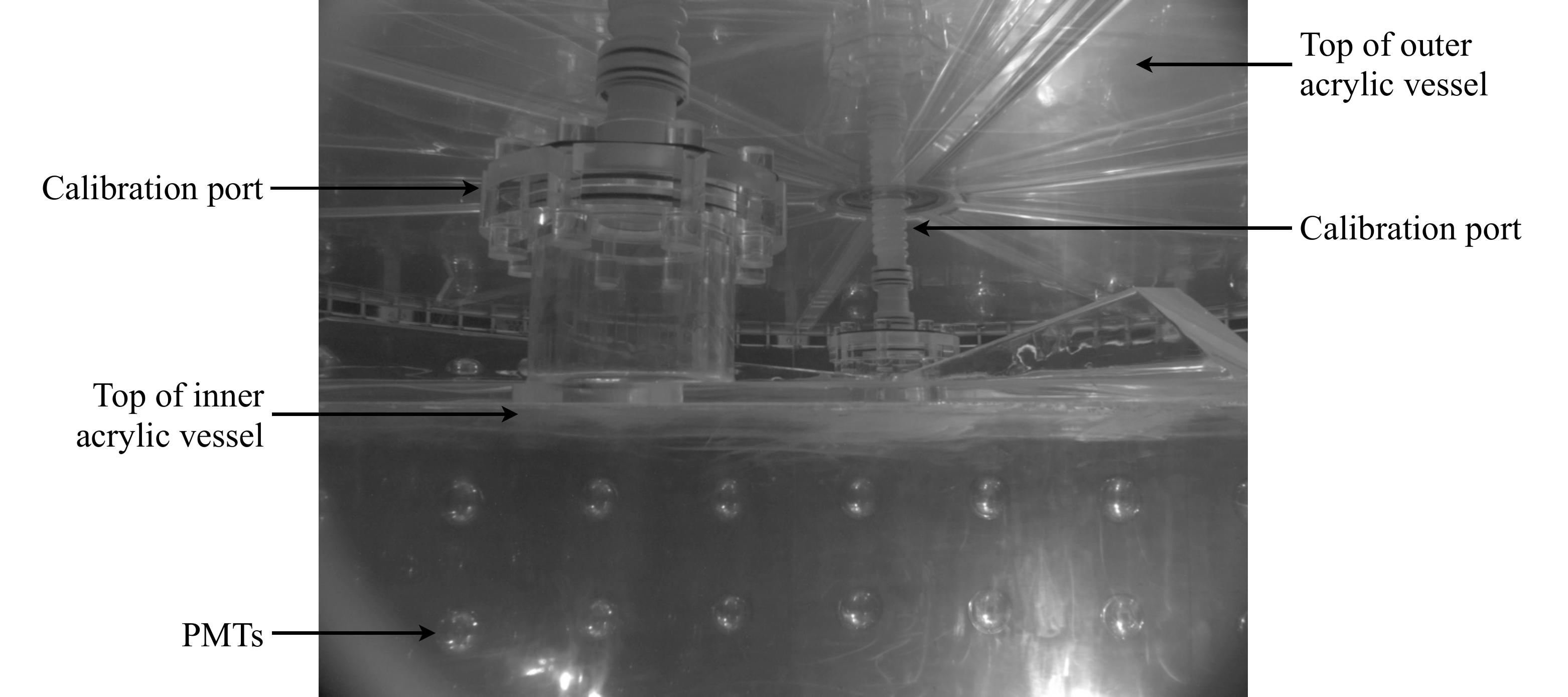}
  \caption{
    View from the top camera in a dry detector.
    Calibration ports and acrylic vessel lids are prominent above the midline of the image.
    Below, several rows of PMTs can be seen. 
    The light source is the white LEDs of the bottom camera.
  }
  \label{fig:topex}
\end{figure}

\begin{figure}[htbp!]
  \centering
  \includegraphics[width=\figwidth]{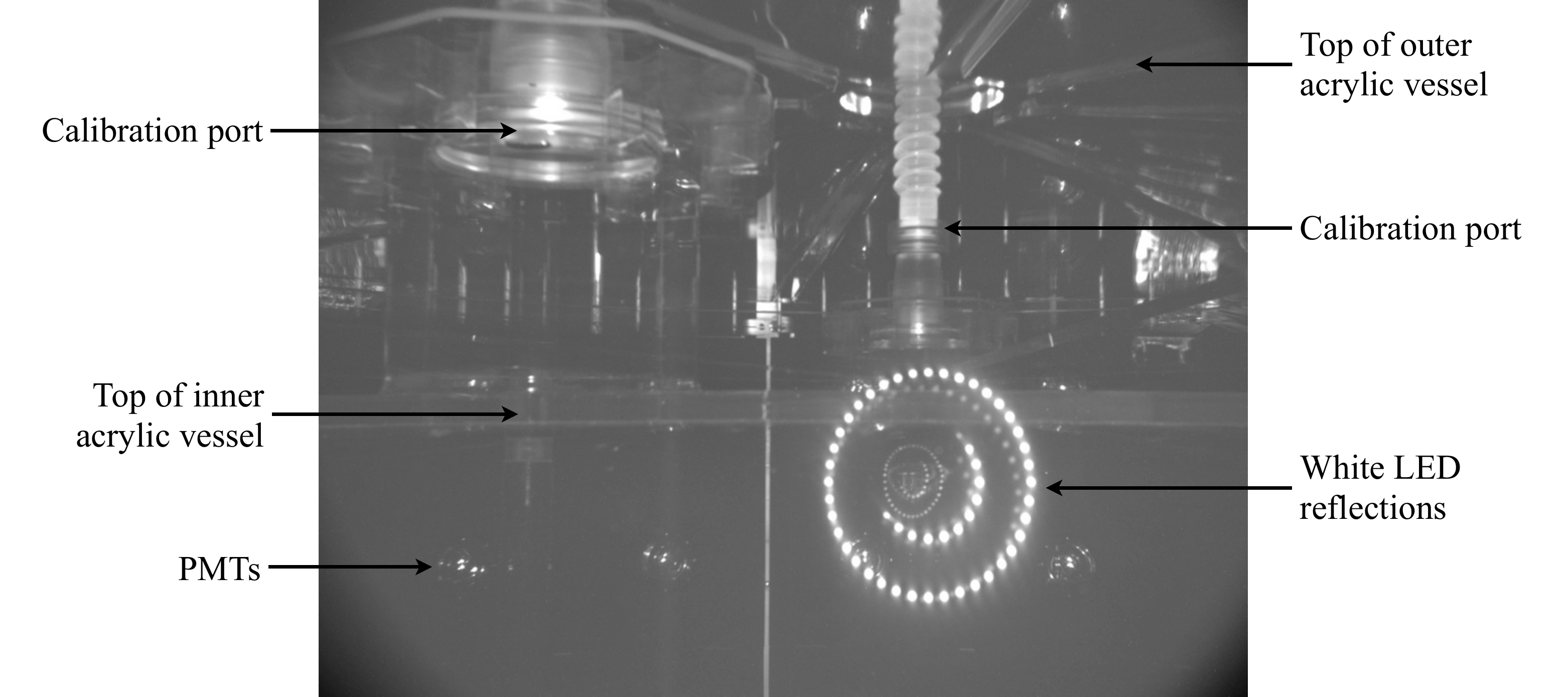}
  \caption{
    Photograph from the top camera after filling completion. 
    The light source is the white LEDs from the same camera. The reflections of the LEDs are visible at 
    the bottom left, where the missing LED reflections are due to the curvature of the acrylic vessels.
    The photograph appears magnified relative to Figure~\protect\ref{fig:topex} because the scintillators have a higher 
    index of refraction than air. 
  }
  \label{fig:filled}
\end{figure}

\begin{figure}[htbp]
  \centering
  \includegraphics[width=\figwidth]{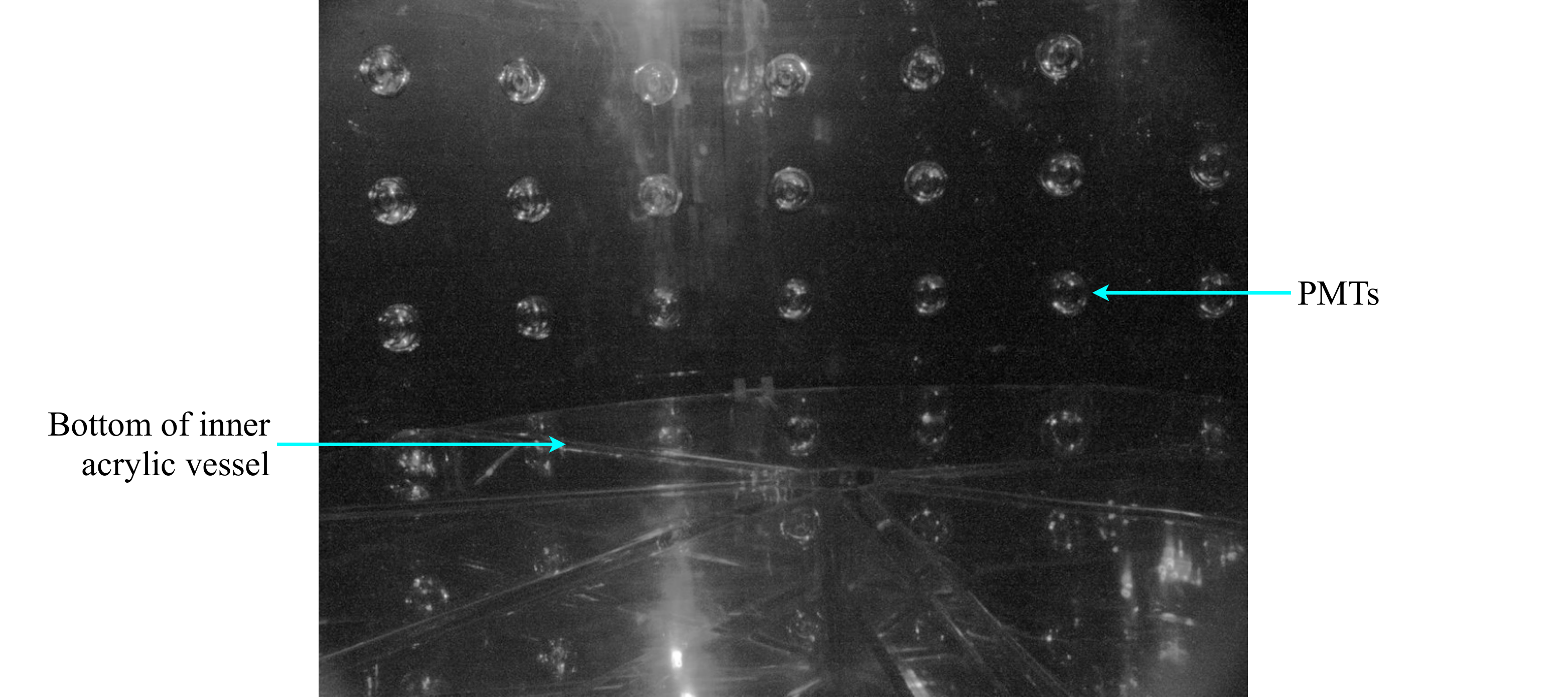}
  \caption{
    Left: A photograph taken with the bottom camera under infrared light from the top camera. 
    The acrylic vessels and PMTs are visible in this detector waiting to be filled.
    The quality is lesser compared to white light, but still effective.
  }
  \label{fig:IRpic}
\end{figure}

\begin{figure}[htbp]
  \centering
  \includegraphics[width=\figwidth]{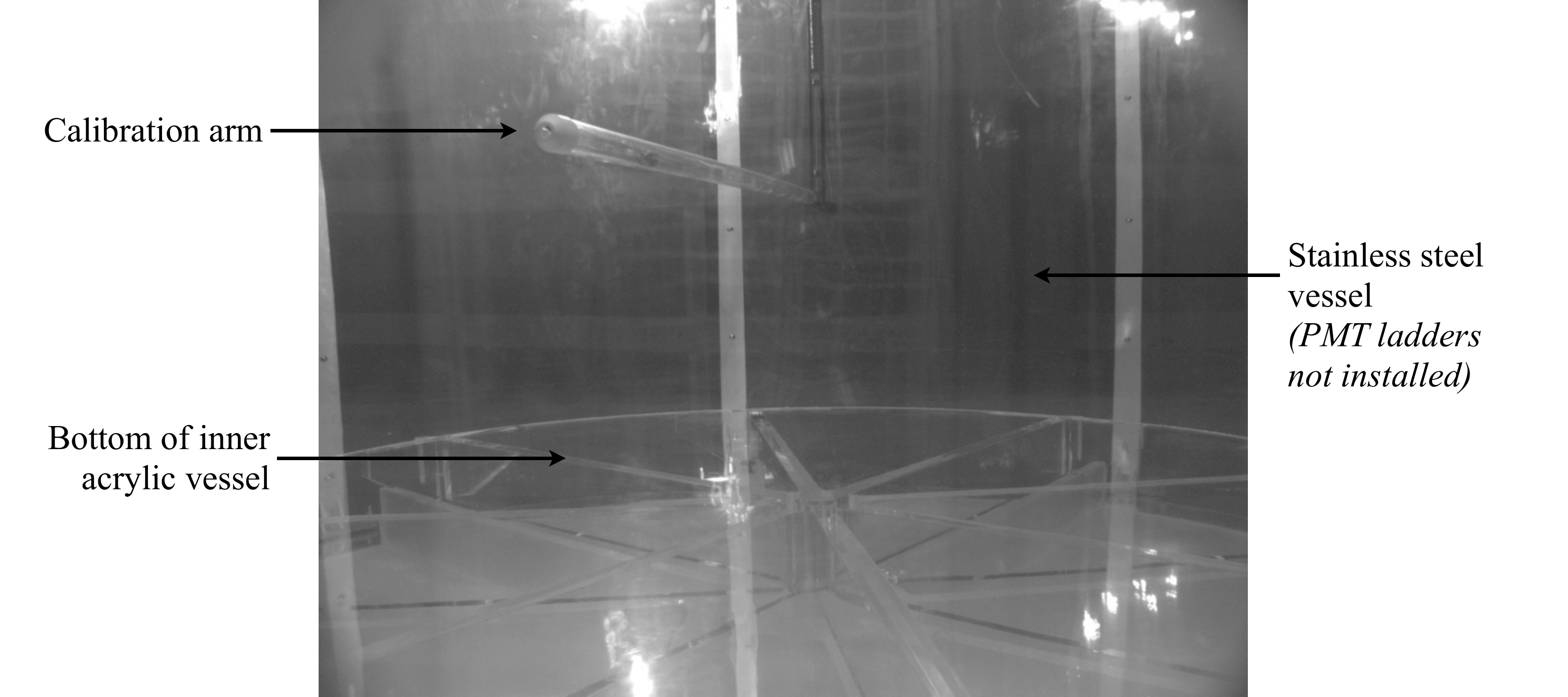}
  \caption{
    A photograph taken of the manual calibration test in a partially 
    constructed detector, under white light from the top camera. 
    The manual calibration arm is seen extended from the center of the detector toward the camera.
    In the background, only the outer acrylic vessel and stainless steel vessel 
    are present; the PMT ladders were installed at the time of this test.
  }
  \label{fig:mancal}
\end{figure}

The cameras have proven very useful and reliable, providing an average of 2,000 images per detector.
Most of these images were taken during the filling process, followed by monitoring of detector transportation.
Figures~\ref{fig:ACU} and~\ref{fig:midfill} 
show annotated photographs taken using the bottom camera and white light from the top camera,
during an automated calibration test, and detector filling respectively. 
The calibration source and liquid levels appear very clearly. 
Figure~\ref{fig:topex} shows a photograph taken from the top camera in a dry detector with white light from
the bottom camera, while Figure~\ref{fig:filled} shows a photograph taken after filling, using white light 
from the same camera. As an example of infrared light photography,
Figure~\ref{fig:IRpic} displays a photograph of the inside of a dry detector taken with the bottom camera
under infrared light from the opposite camera. Finally, Figure~\ref{fig:mancal} shows an image 
of a manual calibration mechanical test in a partially assembled detector, taken under with the bottom camera 
white light from the top camera.

\section{Summary}

The cameras have proven to be a vital component in building and maintaining the antineutrino
detectors for the Daya Bay experiment. All together they have captured over 10,000 photographs
during the filling, transportation and calibration procedures, without interference with data collection. 
The system was effectively designed 
around several constraints imposed by the arrangement of features within the detector, materials 
compatibility, radioactive backgrounds, electronics and light interference. The designs presented in
this article are useful for the design of any camera system or similar low-background application.

\appendix
\section{Acknowledgments}

We gratefully acknowledge support from the Department of Energy 
Office of Science, High Energy Physics, under contract DE-FG02-95ER40896 and support of the
University of Wisconsin Foundation.  
We thank the Kellogg group at the California Institute of Technology and their HPGe radioassay facility,
as well as the Low Background Facility at Lawrence Berkeley National Laboratories,
for measuring radioactive backgrounds of the camera system. 
We thank Jesse Nims and Ruxiu Zhao for assistance in assembling cameras; 
Amy Pagac and Al Riley for providing drawings; Billy J. Gates Jr., Daniel Wahl and Harold Mattison for advice regarding electronics;
Zhimin Wang, Dan Dwyer, Hu Tao, Cheng-Ju Lin, Kam-Biu Luk and Johnny Ho for assistance in testing PMT hit rates 
under infrared light; and Julio De Oliveira for help with evaluating cameras.


\bibliographystyle{JHEP}
\bibliography{admcpaper}

\end{document}